%% file: tom_paper.tex
\documentclass[useAMS,usenatbib]{mn2e}

\usepackage{times}
\usepackage{graphicx}
\usepackage{color}									
\usepackage[hidelinks]{hyperref}					
\usepackage{placeins}
\usepackage[english]{babel}
\usepackage{amsmath}								
\usepackage{longtable}							
\usepackage[flushleft]{threeparttable}			
\usepackage{fixltx2e}                                  

\input{definitions}

\makeatletter
\newcommand*{\rom}[1]{\expandafter\@slowromancap\romannumeral #1@}

\newcommand \Dotfill {\leavevmode \cleaders \hb@xt@ .33em{\hss .\hss }\hfill \kern \z@}
\newcommand\ion[2]{\mbox{#1\hspace{0.2em}{{\footnotesize{\rmfamily\@Roman{#2}}}}}} 
\makeatother

\title[X-ray state transitions of Cygnus X-1]{Interpreting the X-ray state transitions of Cygnus X-1\thanks{In this study we use optical spectra obtained from Ond\v{r}ejov 2.05-m and Xinglong 2.16-m telescopes, and X-ray data from RXTE-ASM, MAXI and {\it{Swift}}-BAT all sky monitors.}}
\author[J. \v{C}echura, S. D. Vrtilek and P. Hadrava]{J. \v{C}echura$^{1,2}$\thanks{E-mail:
cechura@astro.cas.cz (JC); saku@cfa.harvard.edu (SDV); had@asu.cas.cz (PH).}, S. D. Vrtilek$^{2}$\footnotemark[2] and P. Hadrava$^{1}$\footnotemark[2]\\
$^{1}$Astronomical Institute of the Academy of Sciences of the Czech Republic, Bo\v{c}n\'i II 1401/1, Prague, 141 00, Czech Republic\\
$^{2}$Harvard-Smithsonian Center for Astrophysics, 60 Garden Street, Cambridge, MA 02138, USA
}
\begin{document}

\date{Accepted 2015 April 2; Received 2014 December 14}

\pagerange{\pageref{firstpage}--\pageref{lastpage}} \pubyear{2015}

\maketitle

\label{firstpage}

\begin{abstract}
We present a novel method for interpreting observations of high-mass X-ray binaries (HMXBs) based on a combination of spectroscopic data and numerical results from a radiation hydrodynamic model of stellar winds. We calculate synthetic Doppler tomograms of predicted emission in low/hard and high/soft X-ray states and compare them with Doppler tomograms produced using spectra of Cygnus X-1, a prototype of HMXBs. Emission from HMXBs is determined by local conditions within the circumstellar medium, namely density, temperature, and ionization state. These quantities depend strongly on the X-ray state of the systems. By increasing intensity of an X-ray emission produced by the compact companion in the HMXB model, we achieved a complete redistribution of the circumstellar medium in the vicinity of the modelled system. These changes (which simulate the transitions between two major spectral states) are also apparent in the synthetic Doppler tomograms which are in good agreement with the observations. 
\end{abstract}


\section{Introduction}
Cygnus X-1 (V1357 Cyg; HDE 226868) is one of the brightest and widely observed X-ray sources in the sky, and an archetype of high-mass X-ray binaries (HMXBs). The optical component -- an O9.7 Iab supergiant \citep{1986ApJ...304..371G} -- and the compact component -- most likely to be a stellar mass black hole -- revolve around the centre of mass of the system with orbital period 5.6 d. The intense X-ray emission from the compact companion is thought to be powered by the accretion of the circumstellar matter originating from the supergiant. A further period of 294 d \citep{1983ApJ...270..233P}, discovered in both optical and X-ray data, is thought to be associated with precession of the accretion disc. The system undergoes aperiodic transitions among several major X-ray states represented by a global flux and spectral change. The first transition was observed by \cite{1972ApJ...177L...5T}: the soft X-ray flux (2--6 keV) decreased by a factor of 4, while the hard flux (10--20 keV) increased by a factor of 2, and radio emission turned on. Since the X-ray emission is produced through the accretion process of the stellar wind of the donor star on to the compact component it is safe to assume that such a transition is associated with a global redistribution of the circumstellar matter in the system. 

Despite precipitous advances in observational techniques and instruments, the angular resolution of even the largest telescopes is still insufficient to observe the accretion process in HMXBs through direct imaging. We thus have to employ indirect methods such as Doppler tomography which can resolve accretion flows in binaries on micro-arcsecond scales by using time-resolved spectroscopy. This technique utilizes binary spectral line profiles taken at a series of different orbital phases and transforms them into a distribution of emission over the binary. 

In Sec.~\ref{DT5.6}, we give a detailed description of optical spectra that we have available, and the method we used to determine the X-ray state of the system for individual observations. Spectra were obtained in each of the major X-ray states of Cygnus X-1. We analyse the states separately in order to determine any structural changes of the stellar wind during transitions. In Sec.~\ref{DT3.0}, we compare results obtained from the Doppler analysis of the spectra compartmentalized according to the X-ray state. We also discuss influence of length of the spectra-gathering period on the quality of the resulting tomograms. Sec.~\ref{C4SS_IDM} describes input parameters of the radiation hydrodynamic model \citep[referred to below as CH]{2015A&A...575A...5C} of the stellar wind in Cygnus X-1. We use the predicted distribution of the circumstellar matter to create synthetic tomograms for individual X-ray states, and compare them with tomograms produced using H-alpha line.

\section{Observations}
\label{DT5.6}

In order to perform the Doppler mapping of Cygnus X-1, we have collected spectroscopic data from the Perek 2.05-m telescope at Ond\v{r}ejov observatory of the Astronomical Institute of the Academy of Sciences of the Czech Republic. We also included the spectra obtained at 2.16-m telescope at Xinglong station of National Astronomical Observatories, China \citep[NAOC;][]{2008AJ....136..631Y}. The data consist of isolated measurements as well as several series of observations, and cover a period between 2003 and 2013. At the Ond\v{r}ejov observatory, 89 spectra were obtained with CCD in the focal point of 700 mm camera of Coud\'{e} spectrograph. A typical resolution is about 0.25 $\mathrm{\mbox{\r{A}}}$ per pixel. At the Xinglong station, an additional 18 spectra were acquired with a CCD grating spectrograph at Cassegrain focal point of the telescope with an intermediate resolution of 1.22 $\mathrm{\mbox{\r{A}}}$ per pixel. The spectra were obtained during the low/hard, high/soft and intermediate state of Cygnus X-1. The complete list of the spectra used in the Doppler analysis can be found in appendix, including observational date, UT time of the beginning of the observation, exposure time, heliocentric Julian date of the centre of the observation interval, wavelength range, and spectral state. Orbital phase is also indicated, the ephemeris of the inferior conjunction of the companion star is adopted from \cite{2003ApJ...583..424G},
\begin{equation}
	\label{E_DT_02}
	2451730.449 + 5.599829E \ . \nonumber
\end{equation}

The observations were reduced using standard reduction routines in \textsc{iraf}. All spectra were bias-subtracted, flat-field corrected, and had cosmic rays removed. After cleaning and wavelength calibration, the spectra were converted to ASCII format to be further processed in \textsc{molly}, a one-dimensional spectrum analysis program developed by Tom Marsh. In \textsc{molly}, the spectra were corrected for the heliocentric effects, and normalized by fitting polynomials by which the corresponding spectra were divided.

The most prominent feature in the observed spectral range is a variable H$\alpha$ line profile that is strongly dependent on orbital phase and spectral state. It is apparent from the observations that the spectra taken in the low/hard state have a strong emission in the whole H$\alpha$ line profile, while in the high/soft state period, the emission remains only in the long-wavelength wing of the line and the short-wavelength side of the line profiles reveals an absorption, as is typical for the P-Cygni profiles of stars with intense mass-loss via a stellar wind.  

The \ion{He}{1} $\lambda 6678$ line is practically free of emission in both states. It means that this line may enable us to reliably measure radial velocities (RVs) of the primary component to get a constraint on the orbital parameters of the system.
 
One of the pronounced spectral features is the absorption at $\lambda = 6614$ $\mathrm{\mbox{\r{A}}}$. It is actually an example of a diffuse interstellar band (DIB) in our spectra. The carrier molecule of these interstellar absorption bands is still unknown \citep{1995ARA&A..33...19H,2003ssac.proc..147K}. The identification of the carriers of these bands remains an important problem in astronomy to date, and the current consensus on the nature of the carriers is that they are probably carbon-bearing molecules that reside ubiquitously in the interstellar gas \citep{2000ARA&A..38..427E}. The most promising carrier candidates are carbon chains, polycyclic aromatic hydrocarbons (PAHs), and fullerenes \citep{1996ApJ...458..621S,1999ApJ...526..265S,1994Natur.369..296F,2000MNRAS.312..769S,2000ApJ...531..312M}. The DIB at $\lambda = 6614$ $\mathrm{\mbox{\r{A}}}$ has a finer structure consisting of three absorption lines \citep{2002A&A...384..215G} which are, however, below the resolving power of our instruments. Since DIBs do not follow the orbital motion of the binary, we use them to increase accuracy in wavelength scaling. For this purpose, we calculated RVs of the systemic shifts for each spectra using \textsc{korel}, a code employing a method of Fourier disentangling of spectra of binary and multiple stars \citep[cf. e.g.][]{2009arXiv0909.0172H,2012IAUS..282..351H,2008AJ....136..631Y}. 
 
The spectra are of good quality in terms of spectral resolution which makes them suitable for Doppler tomography. However, in the terms of time coverage, things get more complicated. It is assumed in Doppler tomography that the distribution of emission from the system is constant in the course of the whole observational campaign. However, Cygnus X-1 varies also on time-scale shorter than its orbital period $P_\mathrm{orb}\simeq 5.6$ d. The characteristic dynamical time-scale of the circumstellar matter is of the order of $P_\mathrm{orb}$, which is also the minimum duration of observational run needed to cover all orbital phases. Because of this difficulty we thus suppose that, in spite of the short-time variations caused most likely by a turbulent motion of gas, the fundamental distribution of emission remains stable and unique for a particular spectral state. This assumption allows us to divide all observations in groups according to their spectral state. With two exceptions of observational series in 2003 and 2013, most of our data are scattered throughout a period of almost a decade. These observations spread over many temporary fluctuations provide us with a sufficient phase coverage to perform the Doppler tomography and to find the mean distributions in the low/hard and high/soft states. In the case of the observational series in 2003 and 2013, we gathered, in the period of several months, enough spectra for the independent Doppler analysis. Even though the orbital phase coverage is not optimal, we can still make some general conclusions and compare the results with the tomograms obtained from the whole period 2003--2013.

\subsection{Determination of X-ray state of Cygnus X-1}
To determine the X-ray spectral state of Cygnus X-1 we need to use X-ray observational data. In the past decade, state information was readily available using the All Sky Monitor on the Rossi X-ray Timing Explorer \citep[RXTE-ASM]{1996ApJ...469L..33L} and regular pointed monitoring observations. Unfortunately, RXTE ceased science operation on 2012 January 5 (MJD 55931) but even prior to this date since January 2010, ASM experienced an instrumental decline which made the observed data unsuitable for determination of the spectral state. We therefore had to combine X-ray data from multiple sources. Here, various state definitions exist, which use measured count rates or colour \citep{2005AN....326..804R,2008ApJ...678.1237G}, or sophisticated mapping between these measurements and spectral parameters \citep[cf. e.g.][]{2007MNRAS.381..723I,2011MNRAS.416.1324Z}. The former prescription is easy to use, but is very instrument specific and cannot be easily translated to other X-ray all sky monitors. The latter approach requires a sophisticated knowledge of the instrumentation of all sky monitors, as well as of the detailed spectral modelling. Furthermore, the previously used state definitions are slightly inconsistent among themselves. The source behaviour from early 1996 (RXTE launch) until the end of 2004 has been discussed by \cite{2006A&A...447..245W}, who used a crude definition of ASM-based states using only the count rate. This classification is sufficient to distinguish main activity patterns, and also consistent with more detailed studies \citep{2011MNRAS.416.1324Z}. But in this work, we use a novel approach to classify states of Cygnus X-1 using all sky monitors RXTE-ASM, MAXI, and {\it{Swift}}-BAT based on 16 years of pointed RXTE observations and suggested by \cite{2013A&A...554A..88G}. 
\begin{table*} 
  \begin{threeparttable}
	\centering 

	\footnotesize	
	\begin{tabular}{l c c c}
		\hline
		\hline 
State & ASM-based & MAXI-based & BAT-based \\
		\hline
Low/hard & $c\leq 20 \vee c\leq 55(h-h_0)$ & $c_\mathrm{M}\leq 1.4h_\mathrm{M}$ & ... \\
Intermediate & $c > 20 \wedge 55(h-h_0) < c \leq 350(h-h_0)$ & $1.4h_\mathrm{M}<c_\mathrm{M}\leq 8/3h_\mathrm{M}$ & ... \\
High/soft & $c>20 \wedge c> 350(h-h_0)$ & $8/3h_\mathrm{M}<c_\mathrm{M}$ & $c_\mathrm{B}\leq 0.09$ \\
		\hline
	\end{tabular}

  \end{threeparttable}
  	\caption{X-ray state definitions according to different all sky monitors. ASM 1.5--12 keV count rate $c$ is in counts $\mathrm{s}^{-1}$, and $h$ is ASM hardness defined as (5--12 keV/1.5--3 keV), $h_0=0.28$. MAXI 2--4 keV count rate $c_\mathrm{M}$ is in counts $\mathrm{s}^{-1}$, and $h_\mathrm{M}$ is MAXI hardness defined as (4--10 keV/2--4 keV). BAT normalized 15--50 keV count rate $c_\mathrm{B}$ is in counts $\mathrm{cm}^{-2}$ $\mathrm{s}^{-1}$. Distinguishing between hard and intermediate state is not possible from the BAT light curves alone. The source is defined as in BAT-based hard or intermediate state for $c_\mathrm{B}>0.09$ counts $\mathrm{cm}^{-2}$ $\mathrm{s}^{-1}$ \citep{2013A&A...554A..88G}.}
	\label{TAB_DT_02}
\end{table*}

MAXI is an all-sky monitor on board the Japanese module of the International Space Station \citep{2009PASJ...61..999M}. Light curves from the Gas Slit Camera detector (GSC) are available in three energy bands (2--4 keV, 4--10 keV, and 10--20 keV) on a dedicated website\footnote{http://maxi.riken.jp/top/index.php?cid=1\&{}jname=J1958+352}. MAXI light curves show prolonged gaps of several days due to observational constraints. MAXI coverage started on MJD 55058 (2009 August 15).

{\it{Swift}}-BAT is sensitive in the 15--150 keV regime \citep{2005SSRv..120..143B,2013ApJS..209...14K}. Satellite-orbit averaged light curves in the 15--50 keV energy band from this coded mask instrument are available on a dedicated website\footnote{http://swift.gsfc.nasa.gov/results/transients/}. BAT coverage started on MJD 53414 (2005 mid-February).
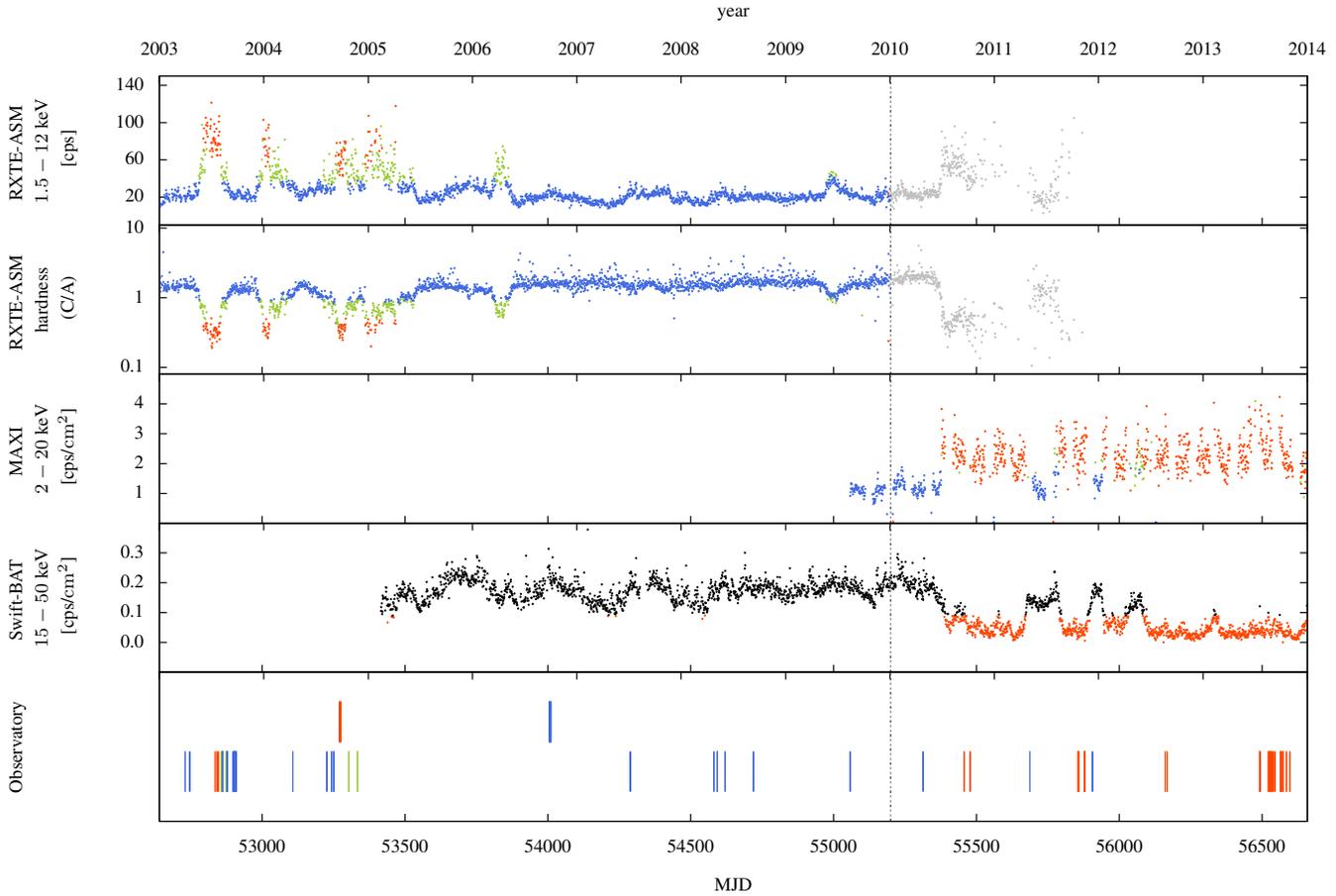
\begin{figure*} 
	\centering
    \resizebox{\textwidth}{!}{\input{DTfig006.tex}}
    \caption{X-ray light curves (RXTE-ASM, MAXI, and {\it{Swift}}-BAT) of Cygnus X-1 in the highest available resolution. ASM hardness is calculated by dividing count rates in $C$ band (5.0--12 keV) by count rates in $A$ band (1.5--3.0 keV). Vertical dotted line represents starting point of instrumental decline of ASM. Blue, green and red colours represent X-ray spectral states of individual measurements classified by respective classification for the given instruments as described in Table~\ref{TAB_DT_02}: blue represents the low/hard, green the intermediate state, and red the high/soft state. ASM data after MJD 55200 (shown in grey) are affected by instrumental decline. The low/hard and intermediate states cannot be distinguished in BAT; data corresponding to these periods of the low/hard or intermediate state are therefore shown in black. The lowest panel shows distribution of the available optical spectra from Ond\v{r}ejov (lower) and  Xinglong (upper). The spectra are coloured according to the spectral state.
    }
    \label{DTfig006}
\end{figure*}    

Table~\ref{TAB_DT_02} shows the determination of state based on the data from different all-sky monitors taken from \citep{2013A&A...554A..88G}. A graphical representation associating light-curve behaviour with spectral state of Cygnus X-1 is given in Fig.~\ref{DTfig006}. The first two panels show the RXTE-ASM data coloured according to their established spectral states (blue is for the low/hard, green for the intermediate and red for the high/soft state). The first panel represents the sum of counts in ASM 1.5--12 keV energy range; the second panel shows the hardness of spectra. The hardness is defined as a ratio of count rates in $C$ band (5.0--12 keV) and $A$ band (1.5--3.0 keV). Due to the deterioration of the ASM following MJD~55200 (depicted by the dotted line) no data from the ASM after this date (grey colour) is used for our analysis. The third and fourth panels show the data from MAXI energy band 2--20 keV and {\it{Swift}}-BAT energy band 15--50 keV, respectively. Black represents the low/hard and intermediate state which are indistinguishable from the {\it{Swift}}-BAT data alone. The lowest panel displays the distribution of optical spectra available for our analysis. The Ond\v{r}ejov spectra lie below while the Xinglong data from 2004 and 2006 are shifted upward. The data are scattered over a period since 2003 to 2013. The best series we have available are 2003 with 35 spectra and 2013 with 20 spectra. The individual observations are coloured according to their corresponding spectral states.
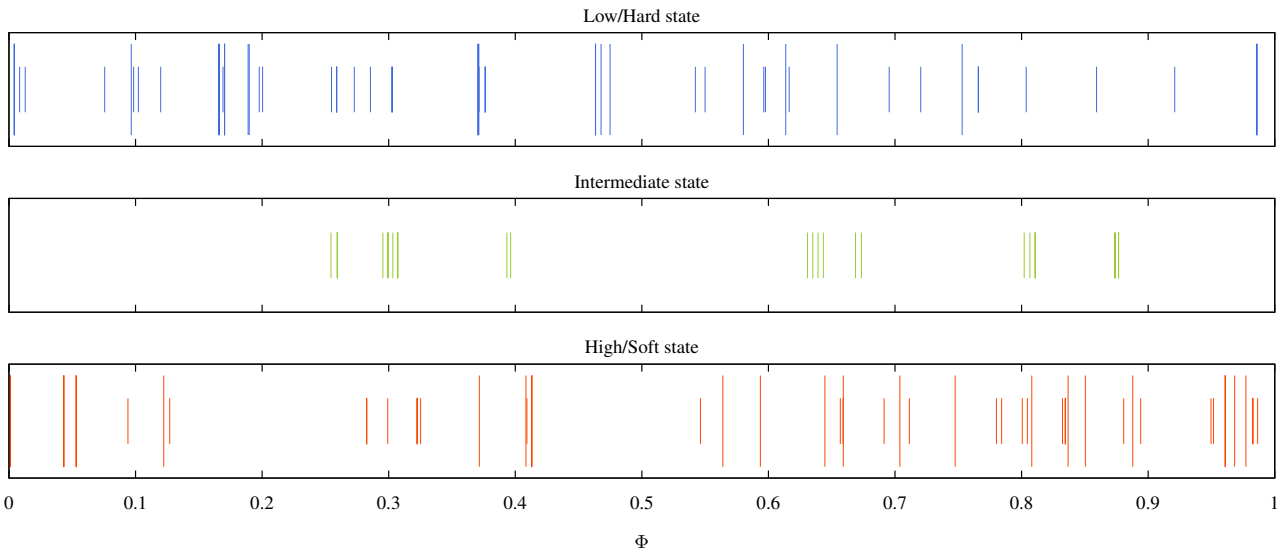
\begin{figure*} 
	\centering
    \resizebox{\textwidth}{!}{\input{DTfig007.tex}}
    \caption{Distribution of the optical spectra of HDE 226868 in orbital phase corresponding to different X-ray states. The low/hard state in blue (44 spectra), the intermediate state in green (19 spectra), and the high/soft state in red (44 spectra). The longer lines in the upper panel represent a distribution of 17 spectra corresponding to the low/hard state in 2003. Similarly in the lower panel, we highlight 20 spectra corresponding to the high/soft state in 2013.
    }
    \label{DTfig007}
\end{figure*}    

In Fig.~\ref{DTfig007}, we show distribution of optical spectra from Ond\v{r}ejov and Xinglong in orbital phase of Cygnus X-1 with respect to their corresponding X-ray states. We assigned 44 spectra to the low/hard state (upper panel), 19 spectra to the intermediate state (middle panel), and 44 remaining spectra to the high/soft state (lower panel). Unfortunately, due to the cadence of our observations, the effective phase coverage is worse than a truly random sample. Despite the clustering problem, the orbital coverage is sufficient to use for Doppler analysis in the low/hard and high/soft states. We highlight (longer lines) 17 spectra taken in 2003 during the low-hard state and 17 spectra taken in 2013 corresponding to the high-soft state.
\begin{figure*} 
	\centering
    \resizebox{\textwidth}{!}{\input{DTfig008.tex}}
    \caption{Optical spectra showing the orbital phase evolution of the major spectral features -- H$\alpha$ line and He \rom{1} $\lambda 6678$. The observation dates and orbital phases are given in the left and right sides of each spectrum, respectively. The left panel shows the selected spectra corresponding to the low/hard state in 2003. The right panel represents the high/soft state in 2013. 
    }
    \label{DTfig008}
\end{figure*}
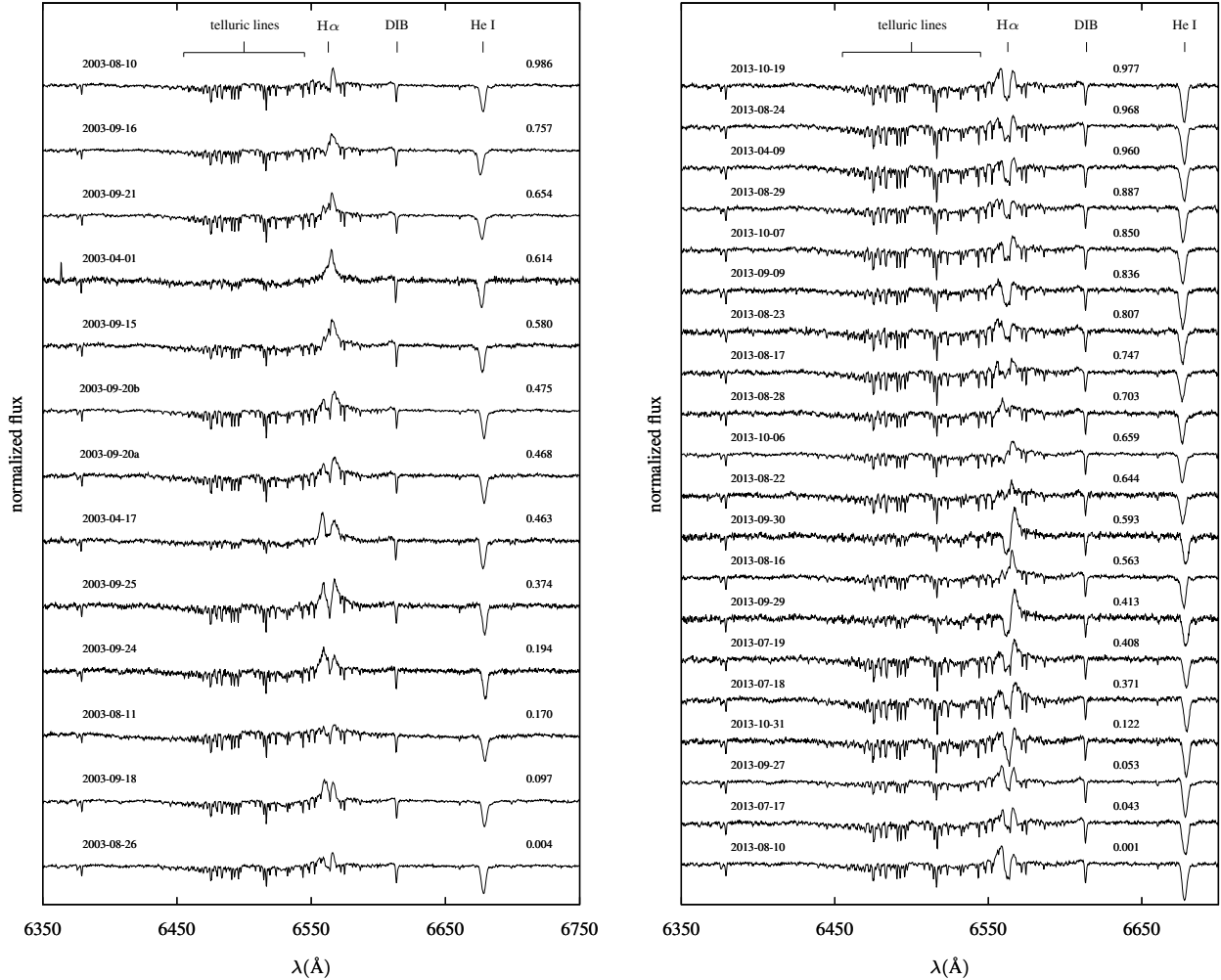    

Fig.~\ref{DTfig008} shows the evolution of the main spectral features of HDE 2266868 over an orbital period. The left panel shows selected optical spectra from 2003 that correspond to the low/hard state of Cygnus X-1. The corresponding observation dates and orbital phases calculated according to the above given ephemeris are written on the left and right side of each spectrum, respectively. Upon close examination, the trailed spectra in H$\alpha$ were found to consist of two superimposed components, shifted in orbital phase. The first component corresponds to a P-Cygni profile of an expanding stellar wind coming from the donor star. The second component is consistent with a presence of strongly variable circumstellar matter in the system. The spectra obtained in 2013, all consistent with the high/soft state, are shown on the right panel. In the high/soft state, only P-Cygni component varying with the orbital movement of the donor is detected. There are no traces of the companion or the circumstellar matter in the \ion{He}{1} $\lambda 6678$ in either spectral state. 

\section{Doppler mapping}
\label{DT3.0}
To analyse our optical spectra, we use \textsc{modmap} -- a program for modulated Doppler tomography using MEM introduced and developed by \cite{2003MNRAS.344..448S}. A grid of pixels spanning velocity space is adjusted to achieve a target goodness of fit measured with $\chi^2$, while simultaneously maximizing the entropy of the image. A refined form of entropy which measures departure from a default image is used. For an excellent summary of the method, see \cite{2001LNP...573....1M}.

For Cygnus X-1, we used the systemic RV $\gamma=-2.7$ km $\mathrm{s}^{-1}$, published by \cite{2006AstL...32..759G}. The systemic velocity needs to be known in order to calculate the correct RV curve of a line source with velocity ($v_x$,$v_y$). If one attempts to reconstruct a data set with an incorrect $\gamma$, it will be difficult to achieve
a good fit to the data since \textsc{modmap} is not tracing the correct velocity curves \citep[cf.][]{1988MNRAS.235..269M}. 

We adopt the orbital period $P_\mathrm{orb}=5.599829\pm 0.000016$ d calculated by \cite{1999A&A...343..861B}. A high precision in the determination of orbital period is needed since we attempt to analyse the phase resolved data spanning over a decade. If the period is not sufficiently well determined, the tomography will fail completely. Less severe consequences would have an incorrect determination of zero-point ephemeris. In this case, the resulted tomogram would be inaccurately positioned. The mass donor star and the compact companion would be rotated by an unknown angle with respect to the emission.

\subsection{The low/hard state}
Here, we present the results of the Doppler tomography of H$\alpha$ $\lambda6562.76$ for the low/hard state of Cygnus X-1. First, we analyse all the spectra (44) corresponding to the low/hard state. Then, we run the analysis on 2003 data only (17 spectra in total) to assess the influence of length of the spectra-gathering period on the quality of the result. For the 2003 data, the distribution of spectra across the orbital phase is not optimal, showing two large gaps around $\Phi=0.29$ and 0.86. However, we still should be able to get reasonable results showing the distribution of emission in the velocity space.  

\begin{figure} 
\centering
  \resizebox{\columnwidth}{!}{\includegraphics{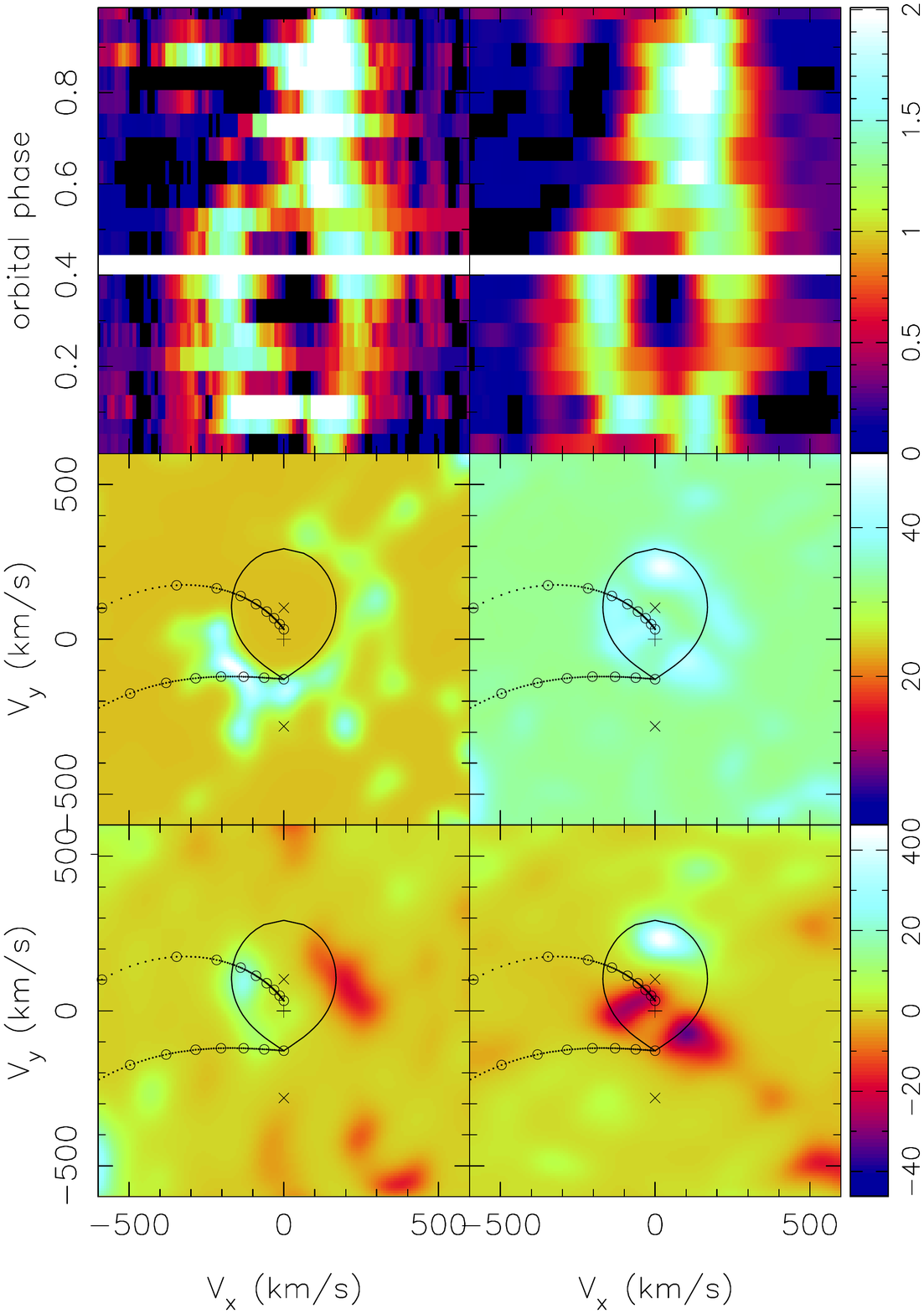}} 
\caption{Modulation Doppler tomogram of H$\alpha$ $\lambda$6562.76 line produced using all spectra from the period 2003--2013 appropriate to the low/hard state of Cygnus X-1. The spectra were divided into 20 phase bins. The figure shows the trailed spectra on top-left panel, the simulated trail on the top-right panel, the distribution of emission in the system in velocity space on the middle-left panel, and the variations of the emission on the middle-right panel. The bottom panels represent the cosine (on the left) and sine (on the right) amplitude maps.
}
\label{DTfig010}
\end{figure}

\begin{figure} 
\centering
  \resizebox{\columnwidth}{!}{\includegraphics{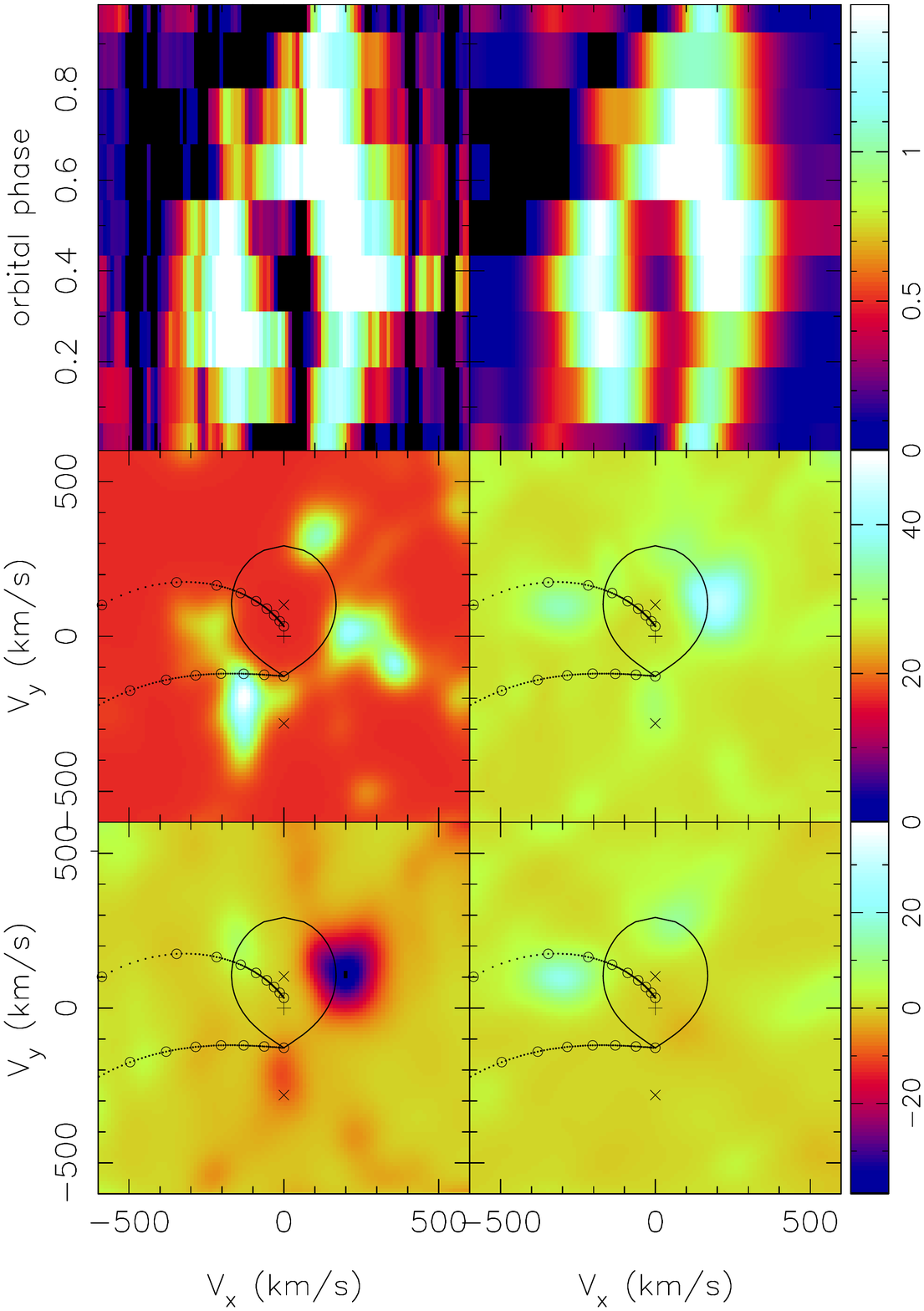}}
\caption{Modulation Doppler tomogram of H$\alpha$ $\lambda$6562.76 line produced using spectra from 2003 only appropriate to the low/hard state of Cygnus X-1. The spectra were divided into 8 phase bins. Panels as defined in Fig.~\ref{DTfig010}.
}
\label{DTfig011}
\end{figure}

The Doppler tomograms of H$\alpha$ $\lambda$6562.76 are shown in Figs~\ref{DTfig010}~and~\ref{DTfig011}. The top panels show the input trailed spectra (on the left) and reproduced trail from the fitted emission distribution (on the right) used to evaluate the goodness of the fit. The four maps in the remaining panels correspond, starting top-left and moving in clockwise fashion; the recovered Doppler image of the average emission from the system, the total modulated emission, the sine amplitude map and the cosine amplitude map of the modulated emission. The spectra in Fig.~\ref{DTfig010} from the period 2003--2013 were binned to the orbital phase. 20 bins were used in total, with only 19 of them were filled. The bin 0.4-0.45 had no appropriate data in it, hence, it remains white. The Doppler tomogram in Fig.~\ref{DTfig011} using data from 2003 only, was produced with spectra divided in eight phase bins -- all of them were filled. 

We see that the reconstructed trail in Figs~\ref{DTfig010}~and~\ref{DTfig011} reproduces the original data well in both cases. Poorer orbital phase resolution in the case the spectra from 2003 is obviously given by the highly limited number of spectra available in this period. The basic structure of the trail, however, remains clear and comparable to the trail constructed from the spectra from the entire observation period 2003--2013. The reconstructed Doppler images in both figures display, similarly, the same general structure. We notice a conspicuous emission spot near the position of the donor star depicted via its critical Roche lobe. The two transferring trajectories are indicated along with the positions of the compact companion and the centre of the donor which are marked with the '$\times$' crosses. The barycentre of the system is shown by a cross '+'. 

\subsection{The high/soft state}
\begin{figure} 
\centering
  \resizebox{\columnwidth}{!}{\includegraphics{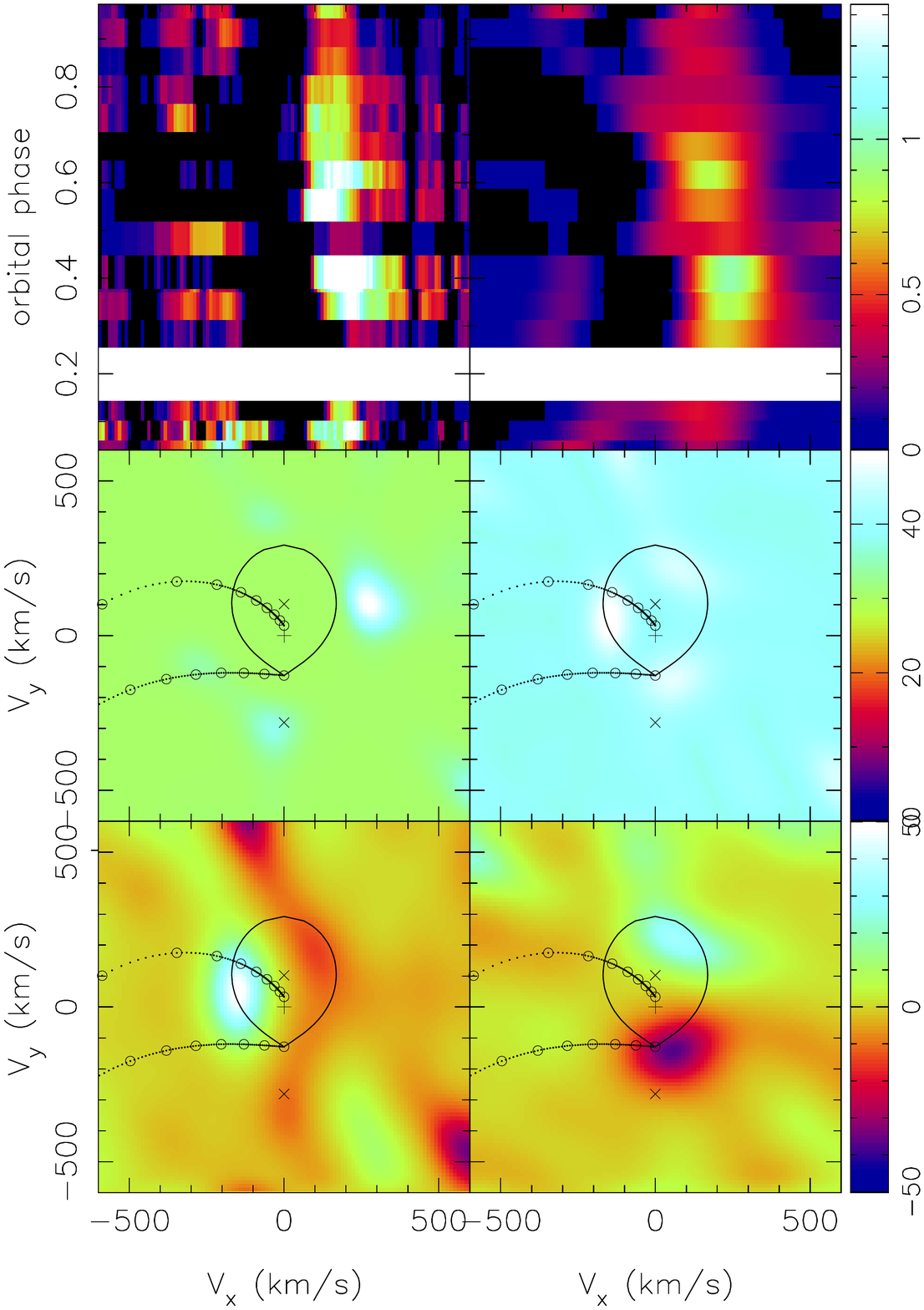}}
\caption{Modulation Doppler tomogram of H$\alpha$ $\lambda$6562.76 line produced using all spectra from the period 2003--2013 appropriate to the high/soft state of Cygnus X-1. The spectra were divided into 17 phase bins. Panels as defined in Fig.~\ref{DTfig010}.
}
\label{DTfig013}
\end{figure}

\begin{figure} 
\centering
  \resizebox{\columnwidth}{!}{\includegraphics{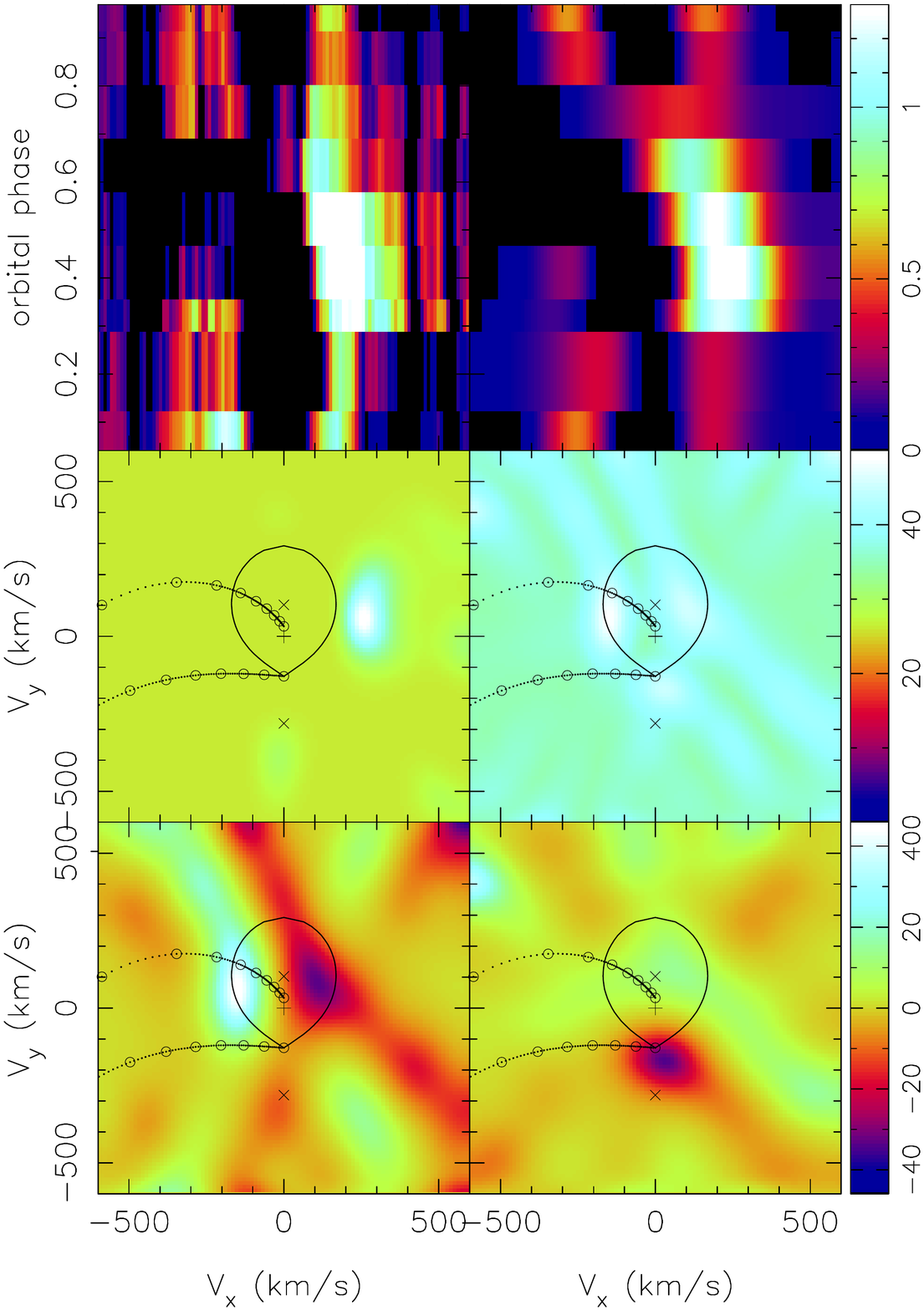}}
\caption{Modulation Doppler tomogram of H$\alpha$ $\lambda$6562.76 line produced using spectra from 2013 when Cygnus X-1 was in the high/soft state. The spectra were divided into nine phase bins. Panels as defined in Fig.~\ref{DTfig010}.
}
\label{DTfig014}
\end{figure}
Following are the results of high/soft state data analysis. First, we apply the Doppler mapping on the entire ensemble of 44 spectra taken in 2003--2013. Then we analyse only the 20 spectra gathered in 2013 over a four-month interval. In this case, the situation is somewhat better in comparison to the low/hard state in 2013. The data, however, also display a couple of large gaps with the most prominent being centred around $\Phi = 0.25$ and 0.5.

The results from modulation mapping of H$\alpha$ line during the high/soft state are shown in Figs~\ref{DTfig013}~and~\ref{DTfig014} for the data covering 2003--2013 period and the 2013 data only, respectively. The spectra in Fig.~\ref{DTfig013} were binned to 17 orbital phase bins, with 16 of them filled and one remained empty. The smaller number of bins is caused with the less ideal orbital phase distribution of the observed data. The Doppler tomogram in Fig.~\ref{DTfig014} uses the 2013 data only, and was produced with spectra decided in nine phase bins -- all of them were filled. 

The reconstructed trails for the high/soft state (Fig.~\ref{DTfig013}~and~\ref{DTfig014}) are not as good as for the low/hard state. This is mainly caused by the larger gaps in the phase distribution of the data and, by extension, by the need of having larger phase bins. Another feature worsening the situation is the complex structure of the H$\alpha$ line trail. This makes the analysis significantly more difficult in the high/soft state. In general, the H$\alpha$ emission is less pronounced than in the low/hard state. But we still notice a number of bright spots in the emission distribution map. These can be remnants of a disc-like structure around the donor star which would correspond to the P-Cygni profile coming from the expanding stellar wind. But at least some of them are more likely to be numerical artefacts. Despite the lesser phase coverage and generally poorer quality of the tomograms in the high/soft state, we can conclude that the emission representing the gas stream from the donor star is no longer there, suggesting substantial redistribution of the circumstellar material during the transition between individual spectral states.

\section{Radiation hydrodynamic model}
\label{C4SS_IDM}
\begin{table}
	\begin{center} 
		\caption{Simulation parameters}
		\label{TAB_01}
		\footnotesize	
		\begin{tabular}{@{}p{4.35cm} l c@{}}
			\hline
			\hline 
			\multicolumn{1}{c}{\rule{0pt}{3ex}Parameter} & \multicolumn{1}{c}{Symbol} & \multicolumn{1}{c}{Used value} \\
			\hline
			\rule{0pt}{3ex}Effective temperature of primary\Dotfill	& $T_\mathrm{eff}$ (K)	& $30000	$\\
			Average radius of primary \Dotfill	& $R_\ast$ ($R_{\odot}$)	& $18$		\\
			Luminosity of primary \Dotfill	& $L_\ast$ ($L_{\odot}$)	& $2.25\times10^5$	\\
			X-ray luminosity & & \\
			\hspace{0.4cm}Low/hard state \Dotfill	& $L_\mathrm{LH}$ ($\mathrm{erg}\ \mathrm{s}^{-1}$)	& $1.9\times10^{37}$	\\
			\hspace{0.4cm}High/soft state \Dotfill	& $L_\mathrm{HS}$ ($\mathrm{erg}\ \mathrm{s}^{-1}$)	& $3.3\times10^{37}$	\\
			Mass of donor \Dotfill	& $M_\ast$ ($M_{\odot}$)	& $24$	\\
			Mass of BH \Dotfill	& $M_\mathrm{x}$ ($M_{\odot}$)	& $8.7$	\\
			Mass-loss rate \Dotfill & $\dot{M}$ ($M_\odot\ \mathrm{yr}^{-1}$) & $2\times10^{-6}$ \\
			Orbital period \Dotfill	& $P_\mathrm{orb}$ (d)	& $5.599829$	\\
			Binary separation \Dotfill	& $D$ ($R_{\odot}$)	& $42.4$	\\
			Eccentricity \Dotfill& $e$ 	& $0$ \\
			\hline
			\label{tab01}
		\end{tabular}
	\end{center}
\end{table}
\begin{figure*} 
\centering
   \setlength{\unitlength}{1mm}
\resizebox{\textwidth}{!}{\input{TP_11.tex}}
\caption{Distributions of physical quantities in the equatorial plane of Cygnus X-1 as calculated with the radiation hydrodynamic model of the stellar wind. The displayed quantities are averaged over 1 $P_\mathrm{orb}$ and represent the density distribution, local temperature, and the ratio of neutral to all hydrogen atoms. The upper and bottom panels correspond to the low/hard and the high/soft state, respectively.} 
\label{TP_11}
\end{figure*}
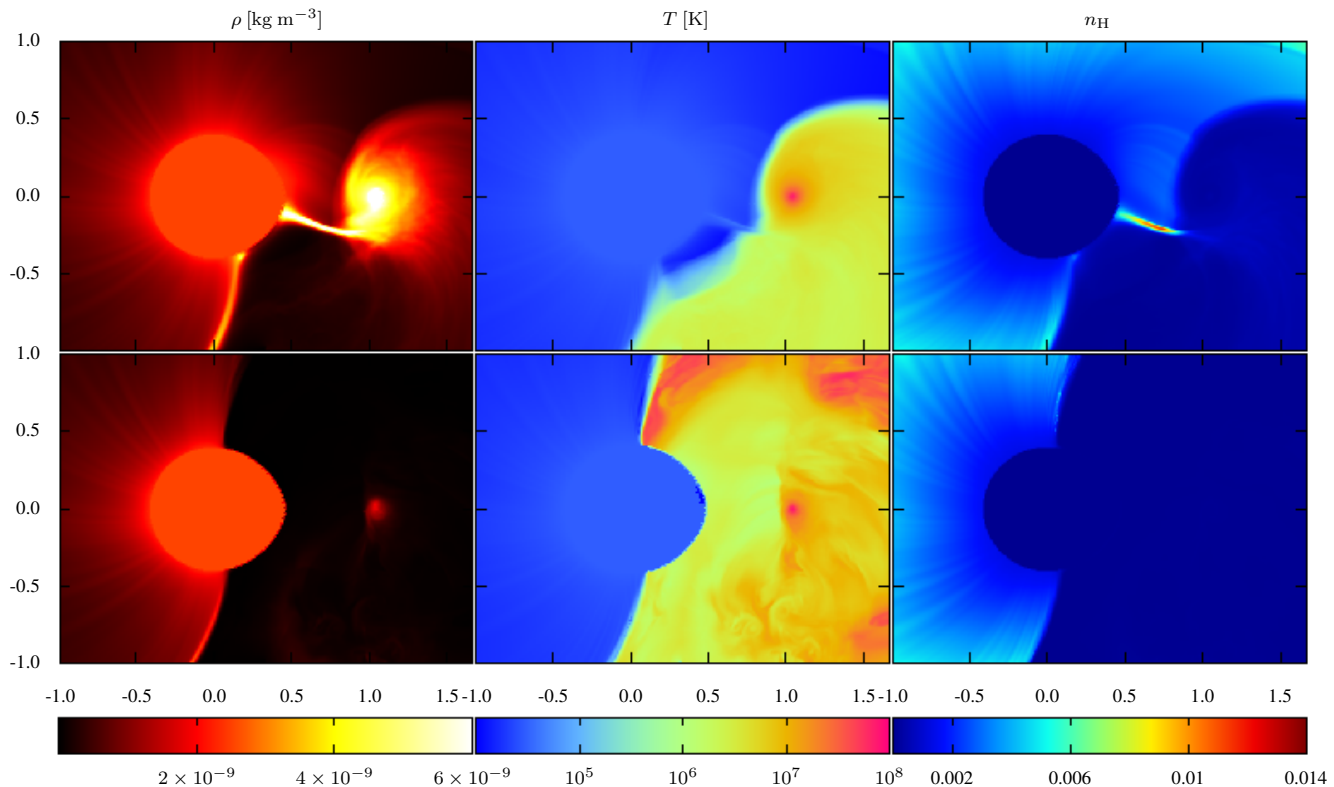
While not as intuitive to interpret as spatial distribution in Cartesian coordinates, Doppler tomograms in velocity space can be obtained without prior assumptions about the velocity field in the flow as a function of position. This significantly simplifies the inversion process and allows application of Doppler tomography in a variety of conditions where the nature of the flow is not a-priori known. A disadvantage is that two spatially very distinct parcels of gas can have precisely the same velocity coordinates; therefore, the transition to spatial coordinates is not uniquely defined.

Aside from this non-uniqueness, it is in principle possible for a given model to convert the velocity coordinates into position coordinates. The only thing that is required is a specification of the velocity at every point in the system. However, there are two complications. First, the transformation between velocity and position is often not known. In fact, it is rarely known, given that it is likely that deviations from Keplerian flow occur. This means that position maps would require re-computation each time system parameters were updated. Secondly, the same place in the system can produce emission at more than one velocity. There are many examples of bright-spot emission  from the gas stream while the disc at the same location produces emission at completely different velocity coordinates \citep[cf. e.g.][]{2015MNRAS.447..149L}. If such data are imaged into position coordinates on the basis of Keplerian rotation, a spot of emission could be produced at a spurious location in the disc. These interpretation difficulties can be potentially resolved only in eclipsing binaries. 

While it is not trivial to convert the tomograms into position coordinates, there is no difficulty in translating predictions of any theoretical model into velocity space. Ideally, the model-observation comparison should be made by predicting trailed spectra, abandoning the Doppler maps altogether. However, Doppler maps still play a role in those theoretical models that are unable to include all peculiarities of real systems, and thus the comparison is easier in velocity coordinates. In CH, we presented an enhanced radiation hydrodynamic model of the stellar wind in HMXBs first introduced by \cite{2012A&A...542A..42H}. We use the model to simulate the structure and dynamics of the circumstellar environment in the vicinity of Cygnus X-1 in both low/hard and high/soft states. Then we convert this information into synthetic Doppler maps directly comparable to the tomograms obtained from the observations.   

Despite being extensively studied for five decades, key parameters of Cygnus X-1 remain uncertain. Many estimates have been made of the masses of both components of the binary -- e.g., \cite{2009ApJ...701.1895C} inferred wide observational limits stating the mass of $23^{+8}_{-6}\ M_{\odot}$ and $11^{+5}_{-3}\ M_{\odot}$ for the primary and the compact component of the system, respectively. However, most of these estimates are unreliable because they are based on unsatisfactory determination of the distance to Cygnus X-1. In their recent paper, \cite{2011ApJ...742...84O} were able to constrain strongly some of the principal parameters of Cygnus X-1. By utilizing an unprecedentedly precise distance from a trigonometric parallax measurement of Cygnus X-1 \citep[$1.86^{+0.12}_{-0.11}\ \mathrm{kpc}$;][]{2011ApJ...742...83R}, they found masses of $M_\ast = 19.16 \pm 1.90\ M_\odot$ and $M_\mathrm{x} = 14.81 \pm 0.98\ M_\odot$ for the O-star and black hole, respectively. These estimates are considerably more robust than previous ones, owing largely to the new parallax distance.

In a recent paper, however, \cite{2014MNRAS.440L..61Z} showed that the mass of the supergiant is inconsistent with the evolutionary models for massive core hydrogen burning stars. Discrepancies of this kind between stellar classification and observed luminosities, inferred masses and temperatures, have previously been reported for the donors in HMXBs \cite[cf., e.g.,][]{1978A&A....63..225C,2001ASSL..264..125K,2015MNRAS.447.1630C}. Based on the evolutionary models, the mass of the supergiant is, most likely, in the range of 25 to 35 $M_\odot$. The corresponding mass of the black hole is in the range of 13 to 23 $M_\odot$. If, as a result of the rotation induced mixing, the hydrogen content of HD 226868 is equal to about 0.6 (as suggested by some observations), then its present mass may be somewhat lower, $\sim 24\ M_\odot$. 

The parameters of the following simulation were chosen in accordance with the observed parameters of Cygnus X-1/HD 226868 (see Table~\ref{TAB_01}). Regarding the above summarized uncertainties, we chose the component masses as a first probe of their values to convey approximately our data at the present time. We fixed the mass of the donor at $M_\ast = 24\ M_\odot$ (a value close to the upper limit of the conceivable mass range) in order to cover a wider range of the mass ratios. Using our optical data, we determined a position of the donor in the velocity space by creating a Doppler map of an inverse \ion{He}{1} absorption line. Because this line originates from the donor itself, such an emission will map to a region that represents the velocity amplitude of the donor. By adjusting the mass ratio of the system, we fitted the size and the shape of the critical Roche lobe on the central non-emitting feature in Figs~\ref{DTfig010} and \ref{DTfig011}, and thus derived the mass of the compact companion equal to $M_\mathrm{x} = 8.7\ M_\odot$. 

The orbital period is 5.6 d and the corresponding orbital separation of the components is $42R_{\odot}$. We assume the primary to be tidally locked and synchronously rotating with the companion orbital motion. See CH for the detailed description of the physical model we use.

In Fig.~\ref{TP_11}, we present the results of our simulations for two values of the X-ray luminosity. The upper panels represent the case with the X-ray luminosity equal to $1.9\ \times\ 10^{37}\ \mathrm{erg}\ \mathrm{s}^{-1}$ corresponding to the low/hard X-ray state of Cygnus X-1. The lower panels show the case of the X-ray luminosity equal to $3.3\ \times\ 10^{37}\ \mathrm{erg}\ \mathrm{s}^{-1}$ which corresponds to the high/soft X-ray state. Each column in Fig.~\ref{TP_11} represents a different physical quantity. From left to right, it is the density distribution $\rho$, the local temperature $T$, and the amount of neutral hydrogen $n_\mathrm{H}$ within the orbital plane of Cygnus X-1. The solution of the model in both cases is quasi-stationary. While the general structure of the flow remains steady and is mainly determined by the X-ray state, the accreting region which is directly affected by the interaction of the shock with the undisturbed wind is rather turbulent and endures quasi-periodic osculations. Since these fluctuations are of a random nature and take place on the time-scales considerably shorter than the orbital period, we filter them out by time-averaging the values of the quantities of interest over 1 $P_\mathrm{orb}$. 

The presence of the strong X-ray source has a profound effect on the structure and dynamics of the wind. Ionizing the gas material severely limits the efficiency of the line-driven mechanism introduced by Castor, Abbott and Klein \citep[hereafter CAK]{1975ApJ...195..157C}, which is the dominant force propelling the wind. In the low/hard state, the outflow from the hemisphere facing the X-ray source is noticeably slowed down. This is in direct contrast with the wind launching from the opposite hemisphere lying in the X-ray shadow cast by the donor where the wind is relatively unaffected by the X-ray radiation. The shadow provides an environment where the wind can be accelerated without interruption, leading to velocities that are an order of magnitude higher than in the slow-wind region where the acceleration mechanism is impaired right at the base of the wind. Slowing down the wind gives a rise to a large accretion disc around the compact object which is preceded by an extensive shock. The stellar wind is highly anisotropic. The gas launched from the hemisphere facing the X-ray source condensates into a narrow dense stream and passes in the vicinity of $L_1$ point in the direction of the compact companion. The mass-loss rate in the direction to the $L_1$ point is thus significantly enhanced. In the high/soft state, the X-ray radiation forms a bubble of full ionization which almost reaches the surface of the donor rendering the CAK mechanism ineffective and preventing the wind to achieve the escape velocity. This effectively cuts off the outflow of the material from the facing hemisphere. The most prominent consequence of the increased X-ray luminosity is the disruption of the narrow dense stream between both components of the binary. The mass transfer from the donor to the accretion disc is then interrupted and the accretion disc shrinks as the matter within it is accreted on to the compact companion. Lacking the incoming flow of gas to compress the shock enveloping the accretion disc, the shock expands in the wide region around the compact companion abruptly increasing the temperature of the remaining gas. Without any prevailing velocity field, the movement of gas in this region is mostly slow and turbulent, and the bow shock around the accretion disc completely disappears. 

\subsection{Production of synthetic Doppler tomograms}
\label{DT5.7}
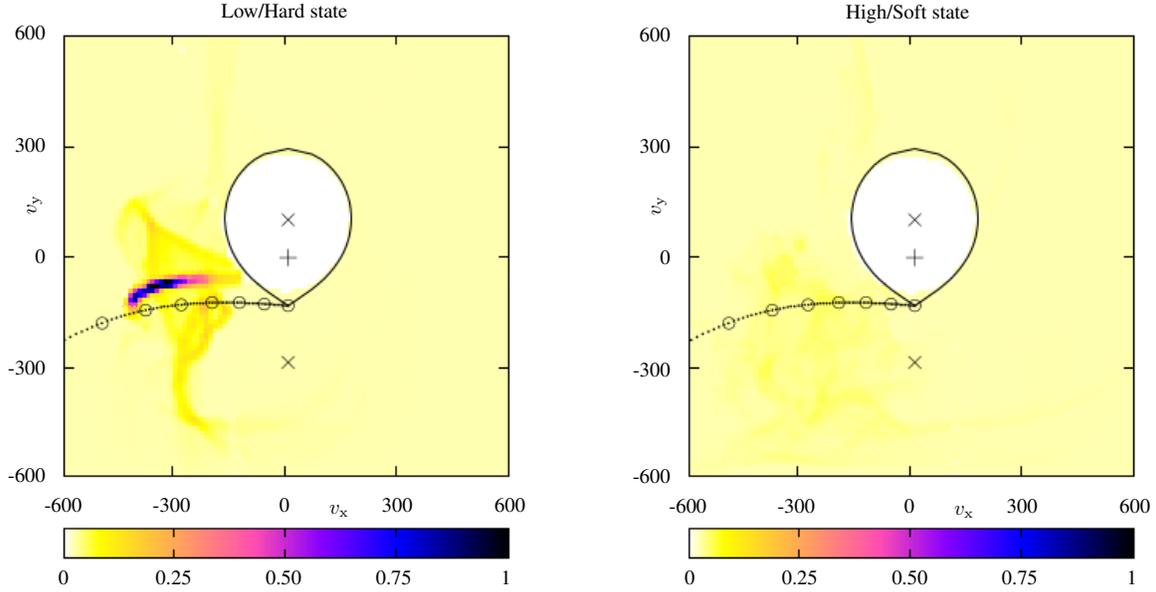
\begin{figure*} 
\centering {\large{
   \setlength{\unitlength}{1mm} 
\resizebox{\textwidth}{!}{\input{TP_12.tex}}}}
\caption{Synthetic Doppler tomograms of the predicted H$\alpha$ line emission based on the model data of neutral hydrogen in two major states of Cygnus X-1 (as shown in the last column of Fig.~\ref{TP_11}). The maps are directly comparable with the Doppler maps acquired from the observational data of HDE 226868 in Fig.~\ref{DTfig010} and \ref{DTfig013}.}
\label{TP_12}
\end{figure*}
\begin{figure*} 
\centering {\large{
   \setlength{\unitlength}{1mm}
\resizebox{\textwidth}{!}{\input{TP_13.tex}}}}
\caption{Associated regions of the wind in spatial and velocity coordinates labelled with corresponding numbers. Displayed quantities are the distribution of neutral hydrogen $n_\mathrm{H}$ in the orbital plane (the left panel) and corresponding relative H$\alpha$ line emission in the velocity space (the right panel).} 
\label{TP_13}
\end{figure*}
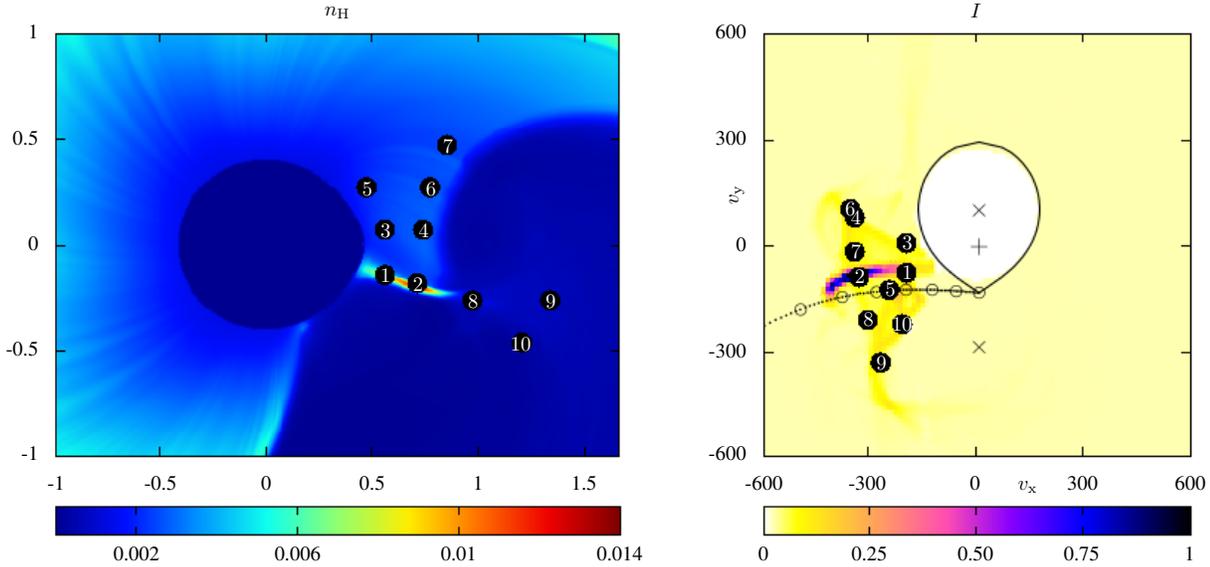
We use the predictions of our model to construct synthetic Doppler tomograms mapping the H$\alpha$ line emission in the low/hard and high/soft state of Cygnus X-1. From the density distribution in the top left panel of Fig.~\ref{TP_11}, we expect that the Doppler maps of Cygnus X-1 in the low/hard state will show the presence of features corresponding to the gas stream and also to the extensive accretion disc. Such a disc would manifest itself in the Dopppler map as a ring-like structure around the position of the compact companion. But we see no signs of the accretion disc present in the spectra of HDE~226868 in Figs~\ref{DTfig010}~and~\ref{DTfig011}. In the high/soft state, we expect that the increased X-ray emission from the compact companion cuts off the gas stream supplying the accretion disc which becomes much less pronounced and possibly not visible at all in H$\alpha$ line. 

To appreciate the conditions within the circumstellar medium in different spectral states, it is convenient to display local temperature and the amount of neutral hydrogen $n_\mathrm{H}$ which we define as a ratio
\begin{equation}
	\label{E_DT_03}
	n_\mathrm{H} = \frac{n(\mathrm{H}^0)}{n_\mathrm{p}+n(\mathrm{H}^0)} \ ,
\end{equation}
where $n(\mathrm{H}^0)$ is number density of neutral hydrogen, and $n_\mathrm{p}$ is proton density. Since an H$\alpha$ line is emitted by neutral hydrogen, $n_\mathrm{H}$ will serve as a tracer of H$\alpha$ line emission. 

We can transform the last column in Fig.~\ref{TP_11} into velocity space. The resultant Doppler maps corresponding to the distribution of neutral hydrogen in both major X-ray states are displayed in Fig.~\ref{TP_12}. The colour-scale shows an expected relative flux in H$\alpha$ line corresponding to the distribution of neutral hydrogen in the orbital plane. We can now directly compare both panels of Fig.~\ref{TP_12} with the Doppler tomograms in Figs~\ref{DTfig010} and \ref{DTfig013} produced from the optical observations. We see that the occurrence of the highest concentration of neutral hydrogen in the velocity space in the synthetic tomogram roughly coincides with the emission region of H$\alpha$ line. As expected, the H$\alpha$ line emission practically disappears when we switched to the high/soft state (right-hand-side panel of Fig.~\ref{TP_12}).  

In Fig.~\ref{TP_11}, the hydrodynamic shock around the accretion region abruptly increases the temperature of the medium up to $10^7$ K. This temperature increase is high enough to cause hydrogen to be completely ionized due to mutual collisions. The region around the compact companion shows only trace amounts of neutral hydrogen leaving the gas stream as the most prominent source of H$\alpha$ line emission in the system. Due to the increased emission of X-ray in the high/soft state (lower panels of Fig.~\ref{TP_11}), the outflow of material from the hemisphere facing the compact companion is interrupted. As a consequence, density in the half-space around the compact companion is considerably diminished, allowing the hydrodynamic shock to expand and heat up the remaining gas. In both cases, however (as shown in the rightmost column of Fig.~\ref{TP_11}), despite the considerable amount of gas there are only traces of neutral hydrogen predicted in the vicinity of the compact companion. Thus, the only strong emission of H$\alpha$ originates in the dense stream which is present in the low/hard state. This emission translates to a Doppler map as a localized maximum of intensity near the ballistic trajectory (left panel of Fig.~\ref{TP_12}). The fact that the maximum does not lie precisely at the ballistic trajectory itself is because such a stream would be launched from the $L_1$ point at the critical Roche equipotential. However, we assume the donor to fill only 95\% of the critical Roche lobe. Thus the stream is shifted towards the centre of primary in $v_y$. The shift towards higher negative $v_x$ is caused by the radiative (CAK-) acceleration which is higher compared to a free-fall of the ballistic trajectory in the effective Roche potential. In the case of the high/soft state which is displayed in the right panel of Fig.~\ref{TP_12}, we notice that the maximum of intensity has disappeared and the remaining emission is scattered and weak. In either case, however, we do not register any contribution of the accretion disc to the H$\alpha$ line emission. 

Besides the expected dominant emission coming from the dense matter flow in the low/hard state, there is also a fine structure of less pronounced ambient emission present in the synthetic Doppler map in the left panel of Fig.~\ref{TP_12}. To get a better idea of which parts of the wind contribute to this emission, we attempt to associate the emission in the velocity space with its spatial position. In Fig.~\ref{TP_13}, we mark some of the regions of the wind and trace them to their corresponding locations in the Doppler tomogram. This approach gives us a set of paired points in velocity and spatial coordinates labelled with numbers. The locations of the points 1 and 2 in velocity space are rather intuitive and coincide with the region of enhanced emission. However other points, although quite distant in spacial terms, map into very similar areas (or even overlap each other) in the velocity space -- e.g. 2--7 or 5--8. This highlights the ambiguity of identifying structures in Doppler tomograms with their corresponding spatial locations. 

\section{Conclusion}
In addition to the radiation hydrodynamic simulations of the stellar wind in HMXBs, we also performed optical observations and Doppler tomographic analysis of the phase-resolved spectroscopic observations of H$\alpha$ line of Cygnus X-1 in the period of 2003--2013. In order to obtain the best orbital phase coverage possible, we put together the spectra we observed (in part) using the Ond\v{r}ejov 2-m telescope in the Czech Republic with the spectra taken at the Xinglong station in China. We utilized the X-ray observations from RXTE, MAXI and {\it{Swift}}-BAT in order to compartmentalize the optical data according to their X-ray spectral state. The results of our analysis show dramatic redistribution of the circumstellar material in Cygnus X-1 during the transition between the low/hard and high/soft states. In the low/hard state, we find an H$\alpha$ emission that is consistent with a dense matter stream (focused stellar wind) in the proximity of the inner Lagrangian point $\mathrm{L}_1$ launching from the donor star and accreting on to the compact companion. Despite the poorer phase coverage leading to worsened quality of the resulted Doppler maps, we conclude that in the high/soft state when the X-ray emission from the compact companion is increased, the H$\alpha$ emission from the matter stream is suppressed. This indicates disruption of the mass transfer between the components of the binary. In the case of the third most prominent X-ray spectral state in Cygnus X-1 -- the intermediate state -- we did not acquire a sufficient amount of data which would allow us to carry out the tomographic analysis. Besides the insufficient phase coverage of the optical spectra, we can expect that during this state the distribution of the circumstellar matter is erratic and highly variable. In order to produce a reliable Doppler tomogram in this X-ray state, we would need observations covering no more than a few orbital periods.  

The results of our simulations suggest (see Fig.~4 in CH) that the large structures, such as the accretion disc or the hydrodynamic shock that envelopes it, are rather insensitive to small variations of the mass ratio of the compact component to the donor. As a consequence, the hydrodynamic simulations are not a suitable tool to better constrain the system parameters such as masses of the individual components. The accretion process, on the other hand, has a decisive influence on the circumstellar matter and the structure of the wind. The X-ray feedback impacts the ionization structure of the wind and thus its ability to be accelerated by the radiation field of the donor. There is also evidence for a jet in Cygnus X-1 \citep{2005Natur.436..819G} which is currently not included in our simulations but which is likely to influence the wind dynamics and may introduce further shocks. Therefore, we plan to further develop the code to enhance its capabilities by adopting an improved physical model. For instance, extending the computation of ionization structure of the wind material to elements other than hydrogen would be very beneficial for our understanding of the wind structure and properties.

The jet of the type observed by \cite{2005Natur.436..819G} would be generated in close proximity to the compact companion. Because of the high computational demands, however, we cannot use the fine resolution necessary to accurately simulate the physical processes in this region. For further advancement in this matter, we have to first improve our model of the accretion disc region taking into account effects of magnetic field and general relativity. This general relativistic radiation magneto-hydrodynamic approach (GR-MHD) would provide us with a more self-consistent model of X-ray binaries. To advance our observation-interpretation method which utilizes the calculated synthetic tomograms, we also intend to acquire data of sources other than Cygnus X-1. For this purpose, low-mass X-ray binaries with typical hours-long orbital periods represent ideal candidates because they allow for full orbital coverage in a matter of one or a few consecutive nights.

Our time-dependent simulations show that even though the wind medium can be susceptible to turbulent motion at the small-scales, the overall solution remains quasi-stationary. This is in agreement with our initial assumption that the large-scale structures present in the wind are stable, and are mainly determined by the X-ray state of Cygnus X-1. Therefore, our use of the optical data that span a period much longer than the orbital period of Cygnus X-1 for Doppler tomography, seems to be well founded. We use the results of our model of Cygnus X-1 to calculate the amount of neutral hydrogen in the circumstellar matter which is responsible for generating the H$\alpha$ line emission. For this distribution, we produce synthetic Doppler tomograms corresponding to  both major X-ray states -- the low/hard and high/soft state -- and compare them with the tomograms obtained from the spectroscopic analysis. The comparison of the observational data with the numerical simulations is done in velocity rather than spatial coordinates to avoid the problems with ambiguity. In both X-ray states, we find the predicted distributions of neutral hydrogen to be in good agreement with the observations, thus confirming the assumed parameters of Cygnus X-1.

\section*{Acknowledgements}
We are grateful to C. S. Peris for helpful suggestions. We also thank an anonymous referee for constructive comments which greatly improved the quality of the paper. Observations by our colleges at Ond\v{r}ejov and Xinlong observatories are highly appreciated. We gratefully acknowledge the use of the \textsc{molly} and \textsc{doppler} software written by T. R. Marsh and the \textsc{modmap} software written by D. Steeghs. This work was supported in part by the Czech Science Foundation (GA\v{C}R 14-37086G) -- Albert Einstein Center for Gravitation and Astrophysics, a grant SVV-260089, and a Smithsonian Institution Endowment Grant awarded to SDV.

\footnotesize{
\bibliographystyle{mn2e}
\bibliography{References}
}
\bsp
\label{lastpage}

\newpage
\onecolumn
\clearpage
\newpage

\appendix

\section[]{}
The complete list of the spectra used in the Doppler analysis is summarized in Table~\ref{TAB_DT_01}.

{\footnotesize{
\begin{longtable}{l c r c c c c}
		\label{TAB_DT_01} \\
		\caption{The list of spectra used for the Doppler tomography}  \\
		\hline
		\hline 
		\multicolumn{1}{c}{Date} & \multicolumn{1}{c}{UT} & \multicolumn{1}{c}{Exposure} & \multicolumn{1}{c}{Heliocentric} & \multicolumn{1}{c}{Wavelength} & \multicolumn{1}{c}{Orbital} & \multicolumn{1}{c}{Spectral} \\
		& \multicolumn{1}{c}{start} & \multicolumn{1}{c}{time} & \multicolumn{1}{c}{Julian date} & \multicolumn{1}{c}{range} & \multicolumn{1}{c}{phase} & \multicolumn{1}{c}{state$^{\ast}$} \\
		\multicolumn{1}{c}{(yyyy/mm/dd)} & \multicolumn{1}{c}{(hh:mm:ss)} & \multicolumn{1}{c}{(s)} & & \multicolumn{1}{c}{($\mathrm{\mbox{\r{A}}}$)} & & \\  
		\hline
		\endfirsthead
		\hline
		\hline 
		\multicolumn{1}{c}{Date} & \multicolumn{1}{c}{UT} & \multicolumn{1}{c}{Exposure} & \multicolumn{1}{c}{Heliocentric} & \multicolumn{1}{c}{Wavelength} & \multicolumn{1}{c}{Orbital} & \multicolumn{1}{c}{Spectral} \\
		& \multicolumn{1}{c}{start} & \multicolumn{1}{c}{time} & \multicolumn{1}{c}{Julian date} & \multicolumn{1}{c}{range} & \multicolumn{1}{c}{phase} & \multicolumn{1}{c}{state$^{\ast}$} \\
		\multicolumn{1}{c}{(yyyy/mm/dd)} & \multicolumn{1}{c}{(hh:mm:ss)} & \multicolumn{1}{c}{(s)} & & \multicolumn{1}{c}{($\mathrm{\mbox{\r{A}}}$)} & & \\ 
		\hline		
		\endhead
		\hline		
		\endfoot
		\
		\endlastfoot
 2003/04/01 & 03:27:09 & 2000 & 2452730.6420 & 6257-6769 & 0.613 & LH \\
 2003/04/17 & 02:14:28 & 3466 & 2452746.5924 & 6258-6770 & 0.463 & LH \\
 2003/07/14 & 21:55:29 & 4000 & 2452835.4167 & 6258-6771 & 0.326 & HS \\
 2003/07/20 & 23:39:47 & 3600 & 2452841.4893 & 6258-6771 & 0.231 & HS \\
 2003/07/25 & 23:16:59 & 3600 & 2452846.4735 & 6258-6771 & 0.300 & HS \\
 2003/08/05 & 22:28:34 &  600 & 2452857.4399 & 6258-6771 & 0.255 & I \\
 2003/08/05 & 22:43:14 & 3600 & 2452857.4501 & 6258-6771 & 0.260 & I \\
 2003/08/08 & 23:54:02 & 1800 & 2452860.4993 & 6259-6771 & 0.624 & I \\
 2003/08/09 & 00:30:30 & 1800 & 2452860.5246 & 6259-6771 & 0.807 & I \\
 2003/08/09 & 01:04:10 & 1800 & 2452860.5480 & 6259-6771 & 0.811 & I \\
 2003/08/10 & 00:21:41 & 3600 & 2452861.5185 & 6258-6771 & 0.987 & LH \\
 2003/08/11 & 01:01:03 &  300 & 2452862.5458 & 6259-6771 & 0.167 & LH \\
 2003/08/11 & 01:08:55 & 3600 & 2452862.5513 & 6259-6771 & 0.171 & LH \\
 2003/08/22 & 22:59:45 & 1800 & 2452874.4614 & 6259-6772 & 0.296 & I \\
 2003/08/22 & 23:32:08 & 1800 & 2452874.4839 & 6259-6772 & 0.300 & I \\
 2003/08/23 & 00:03:43 & 1800 & 2452874.5059 & 6259-6772 & 0.304 & I \\
 2003/08/23 & 00:34:51 & 1800 & 2452874.5275 & 6259-6772 & 0.308 & I \\
 2003/08/24 & 20:20:45 & 1800 & 2452876.3510 & 6259-6772 & 0.632 & I \\
 2003/08/24 & 20:54:48 & 1800 & 2452876.3746 & 6259-6772 & 0.636 & I \\
 2003/08/24 & 21:27:09 & 1800 & 2452876.3971 & 6259-6772 & 0.640 & I \\
 2003/08/24 & 22:01:29 & 1800 & 2452876.4209 & 6259-6772 & 0.644 & I \\
 2003/08/25 & 01:26:12 & 1800 & 2452876.5631 & 6259-6772 & 0.669 & I \\
 2003/08/25 & 02:04:27 & 1800 & 2452876.5897 & 6259-6772 & 0.674 & I \\
 2003/08/26 & 21:52:08 &  400 & 2452878.4144 & 6259-6772 & 0.000 & LH \\
 2003/08/26 & 22:02:47 & 3300 & 2452878.4218 & 6259-6772 & 0.005 & LH \\
 2003/09/15 & 22:33:24 & 4000 & 2452898.4425 & 6258-6770 & 0.581 & LH \\
 2003/09/16 & 22:20:30 & 3600 & 2452899.4335 & 6258-6770 & 0.754 & LH \\
 2003/09/18 & 19:57:43 & 4000 & 2452901.3343 & 6261-6773 & 0.097 & LH \\
 2003/09/20 & 22:13:42 & 1200 & 2452903.4286 & 6259-6772 & 0.468 & LH \\
 2003/09/20 & 22:37:49 & 5175 & 2452903.4454 & 6259-6772 & 0.475 & LH \\
 2003/09/21 & 22:41:52 & 5499 & 2452904.4481 & 6259-6771 & 0.655 & LH \\
 2003/09/24 & 23:19:04 &  300 & 2452907.4756 & 6258-6770 & 0.190 & LH \\
 2003/09/24 & 23:26:23 & 3600 & 2452907.4998 & 6258-6770 & 0.194 & LH \\
 2003/09/25 & 23:44:19 &  180 & 2452908.4924 & 6258-6770 & 0.371 & LH \\
 2003/09/25 & 23:49:54 & 3077 & 2452908.5130 & 6258-6770 & 0.375 & LH \\
 2004/04/12 & 00:33:19 & 7200 & 2453107.5636 & 6257-6769 & 0.921 & LH \\
 2004/08/09 & 00:20:20 & 6300 & 2453226.5540 & 6264-6776 & 0.170 & LH \\
 2004/08/24 & 22:46:10 &  300 & 2453242.4537 & 6263-6776 & 0.009 & LH \\
 2004/08/24 & 22:54:29 & 3600 & 2453242.4786 & 6263-6776 & 0.013 & LH \\
 2004/09/02 & 20:13:36 &  300 & 2453251.3475 & 6263-6776 & 0.597 & LH \\
 2004/09/02 & 20:24:29 & 1800 & 2453251.3638 & 6263-6776 & 0.600 & LH \\
 2004/09/21$^{\ast\ast}$ & 14:46:04 &  900 & 2453270.1205 & 6338-6718 & 0.949 & HS \\
 2004/09/21$^{\ast\ast}$ & 15:01:51 &  900 & 2453270.1315 & 6338-6718 & 0.951 & HS \\
 2004/09/22$^{\ast\ast}$ & 14:36:11 &  900 & 2453271.1137 & 6337-6811 & 0.126 & HS \\
 2004/09/25$^{\ast\ast}$ & 13:47:11 & 1000 & 2453274.0802 & 6338-6738 & 0.656 & HS \\
 2004/09/25$^{\ast\ast}$ & 14:04:44 & 1000 & 2453274.0924 & 6338-6738 & 0.658 & HS \\
 2004/09/26$^{\ast\ast}$ & 13:23:08 & 1000 & 2453275.0635 & 6340-6700 & 0.831 & HS \\
 2004/09/26$^{\ast\ast}$ & 13:40:42 & 1000 & 2453275.0757 & 6340-6700 & 0.834 & HS \\
 2004/10/24 & 19:01:57 &  300 & 2453303.2954 & 6266-6779 & 0.874 & I \\
 2004/10/24 & 19:12:04 & 1800 & 2453303.3112 & 6266-6779 & 0.877 & I \\
 2004/11/24 & 16:50:00 &  300 & 2453334.2020 & 6264-6776 & 0.393 & I \\
 2004/11/24 & 16:57:04 & 2400 & 2453334.2191 & 6264-6776 & 0.396 & I \\
 2006/09/26$^{\ast\ast}$ & 13:55:43 & 1200 & 2454005.0873 & 6339-6777 & 0.196 & LH \\
 2006/09/26$^{\ast\ast}$ & 14:17:18 & 1200 & 2454005.1023 & 6339-6777 & 0.199 & LH \\
 2006/09/27$^{\ast\ast}$ & 13:16:20 & 1200 & 2454006.0600 & 6339-6777 & 0.370 & LH \\
 2006/09/27$^{\ast\ast}$ & 13:55:07 & 1200 & 2454006.0869 & 6339-6777 & 0.375 & LH \\
 2006/09/28$^{\ast\ast}$ & 12:14:02 & 1200 & 2454007.0167 & 6339-6866 & 0.541 & LH \\
 2006/09/28$^{\ast\ast}$ & 13:15:46 & 1200 & 2454007.0596 & 6340-6867 & 0.549 & LH \\
 2006/09/29$^{\ast\ast}$ & 12:19:54 & 1200 & 2454008.0138 & 6338-6787 & 0.720 & LH \\
 2006/10/01$^{\ast\ast}$ & 11:59:59 &  600 & 2454010.0035 & 6335-6798 & 0.075 & LH \\
 2006/10/01$^{\ast\ast}$ & 15:04:09 &  600 & 2454010.1314 & 6334-6797 & 0.098 & LH \\
 2006/10/02$^{\ast\ast}$ & 12:04:59 &  600 & 2454011.0069 & 6333-6797 & 0.254 & LH \\
 2006/10/02$^{\ast\ast}$ & 14:30:00 &  600 & 2454011.1076 & 6332-6796 & 0.272 & LH \\
 2007/07/06 & 22:49:27 & 3600 & 2454288.4748 & 6257-6769 & 0.804 & LH \\
 2008/04/24 & 21:24:46 & 7200 & 2454581.4334 & 6251-6764 & 0.120 & LH \\
 2008/05/06 & 20:53:11 & 7200 & 2454593.4121 & 6251-6764 & 0.259 & LH \\
 2008/05/07 & 00:28:20 & 7200 & 2454593.5615 & 6251-6764 & 0.286 & LH \\
 2008/06/02 & 23:48:45 & 7200 & 2454620.5355 & 6249-6762 & 0.103 & LH \\
 2008/09/09 & 21:36:53 & 8000 & 2454719.4497 & 6252-6765 & 0.766 & LH \\
 2009/08/14 & 22:07:44 & 3600 & 2455058.4462 & 6253-6765 & 0.303 & LH \\
 2010/04/27 & 00:53:04 & 2600 & 2455313.5515 & 6261-6774 & 0.859 & LH \\
 2010/09/17 & 20:53:54 & 4200 & 2455457.3976 & 6256-6768 & 0.547 & HS \\
 2010/10/08 & 19:07:35 & 3600 & 2455478.3194 & 6255-6767 & 0.283 & HS \\
 2011/05/05 & 20:52:34 & 1836 & 2455687.3805 & 6252-6765 & 0.616 & LH \\
 2011/10/21 & 18:45:58 & 1800 & 2455856.2932 & 6252-6764 & 0.780 & HS \\
 2011/10/21 & 19:19:18 & 1800 & 2455856.3164 & 6252-6764 & 0.784 & HS \\
 2011/10/21 & 21:30:30 & 1800 & 2455856.4075 & 6252-6764 & 0.801 & HS \\
 2011/10/21 & 22:03:19 & 1800 & 2455856.4303 & 6252-6764 & 0.805 & HS \\
 2011/10/22 & 21:59:24 & 1800 & 2455857.4275 & 6251-6763 & 0.983 & HS \\
 2011/10/22 & 22:32:19 & 1500 & 2455857.4486 & 6251-6763 & 0.987 & HS \\
 2011/11/12 & 16:25:29 & 1800 & 2455878.1944 & 6252-6764 & 0.691 & HS \\
 2011/11/12 & 19:06:26 & 1800 & 2455878.3062 & 6252-6764 & 0.711 & HS \\
 2011/11/13 & 17:51:39 & 1800 & 2455879.2542 & 6252-6765 & 0.881 & HS \\
 2011/11/13 & 19:40:10 & 1800 & 2455879.3295 & 6252-6765 & 0.894 & HS \\
 2011/12/10 & 16:56:27 & 1800 & 2455906.2144 & 6252-6765 & 0.695 & LH \\
 2012/08/20 & 22:21:20 & 1031 & 2456160.4408 & 6253-6766 & 0.094 & HS \\
 2012/08/27 & 18:58:18 & 4500 & 2456167.3197 & 6253-6766 & 0.322 & HS \\
 2013/07/17 & 00:47:50 & 1801 & 2456490.5469 & 6263-6736 & 0.043 & HS \\
 2013/07/18 & 20:52:57 & 2001 & 2456492.3850 & 6263-6736 & 0.372 & HS \\
 2013/07/19 & 01:55:16 & 1394 & 2456492.5914 & 6263-6736 & 0.408 & HS \\
 2013/08/16 & 22:35:10 & 3001 & 2456521.4618 & 6262-6735 & 0.564 & HS \\
 2013/08/17 & 23:15:07 & 3001 & 2456522.4895 & 6262-6735 & 0.748 & HS \\
 2013/08/22 & 23:48:31 & 3001 & 2456527.5127 & 6262-6735 & 0.645 & HS \\
 2013/08/23 & 21:20:18 & 6203 & 2456528.4283 & 6262-6735 & 0.808 & HS \\
 2013/08/24 & 19:00:15 & 5256 & 2456529.3255 & 6262-6734 & 0.968 & HS \\
 2013/08/28 & 22:16:02 & 2307 & 2456533.4443 & 6262-6734 & 0.704 & HS \\
 2013/08/29 & 22:53:50 & 3001 & 2456534.4746 & 6262-6734 & 0.888 & HS \\
 2013/09/04 & 23:07:07 & 3001 & 2456540.4836 & 6262-6734 & 0.961 & HS \\
 2013/09/09 & 20:52:24 & 2701 & 2456545.3882 & 6262-6734 & 0.837 & HS \\
 2013/09/27 & 20:17:04 & 8991 & 2456563.3993 & 6262-6735 & 0.053 & HS \\
 2013/09/29 & 21:34:03 & 2501 & 2456565.4151 & 6262-6735 & 0.413 & HS \\
 2013/09/30 & 21:40:51 & 3601 & 2456566.4262 & 6262-6734 & 0.594 & HS \\
 2013/10/06 & 20:30:59 & 6282 & 2456572.3929 & 6262-6734 & 0.659 & HS \\
 2013/10/07 & 22:23:27 & 5114 & 2456573.4642 & 6262-6734 & 0.850 & HS \\
 2013/10/08 & 18:55:12 & 2968 & 2456574.3071 & 6262-6735 & 0.001 & HS \\
 2013/10/19 & 20:23:17 & 4037 & 2456585.3739 & 6262-6735 & 0.977 & HS \\
 2013/10/31 & 20:43:25 & 3795 & 2456597.3858 & 6262-6735 & 0.122 & HS \\
 \hline
 \multicolumn{7}{l}{$^{\ast\ }$ low/hard state (LH), high/soft state (HS) and intermediate state (I)} \\
 \multicolumn{7}{l}{$^{\ast\ast}$ spectra observed at the Xinglong station (China)}
\end{longtable}
}}
\twocolumn

\end{document}

%% file: definitions.tex
\def\aj{AJ}%
\def\araa{ARA\&A}%
\def\apj{ApJ}%
\def\apjl{ApJ}%
\def\apjs{ApJS}%
\def\aap{A\&A}%
\def\mnras{MNRAS}%
\def\pasj{PASJ}%
\def\ssr{Space~Sci.~Rev.}%
\def\nat{Nature}%

%% file: DTfig006.tex
\begingroup
  \makeatletter
  \providecommand\color[2][]{%
    \GenericError{(gnuplot) \space\space\space\@spaces}{%
      Package color not loaded in conjunction with
      terminal option `colourtext'%
    }{See the gnuplot documentation for explanation.%
    }{Either use 'blacktext' in gnuplot or load the package
      color.sty in LaTeX.}%
    \renewcommand\color[2][]{}%
  }%
  \providecommand\includegraphics[2][]{%
    \GenericError{(gnuplot) \space\space\space\@spaces}{%
      Package graphicx or graphics not loaded%
    }{See the gnuplot documentation for explanation.%
    }{The gnuplot epslatex terminal needs graphicx.sty or graphics.sty.}%
    \renewcommand\includegraphics[2][]{}%
  }%
  \providecommand\rotatebox[2]{#2}%
  \@ifundefined{ifGPcolor}{%
    \newif\ifGPcolor
    \GPcolortrue
  }{}%
  \@ifundefined{ifGPblacktext}{%
    \newif\ifGPblacktext
    \GPblacktextfalse
  }{}%
  \let\gplgaddtomacro\g@addto@macro
  \gdef\gplbacktext{}%
  \gdef\gplfronttext{}%
  \makeatother
  \ifGPblacktext
    \def\colorrgb#1{}%
    \def\colorgray#1{}%
  \else
    \ifGPcolor
      \def\colorrgb#1{\color[rgb]{#1}}%
      \def\colorgray#1{\color[gray]{#1}}%
      \expandafter\def\csname LTw\endcsname{\color{white}}%
      \expandafter\def\csname LTb\endcsname{\color{black}}%
      \expandafter\def\csname LTa\endcsname{\color{black}}%
      \expandafter\def\csname LT0\endcsname{\color[rgb]{1,0,0}}%
      \expandafter\def\csname LT1\endcsname{\color[rgb]{0,1,0}}%
      \expandafter\def\csname LT2\endcsname{\color[rgb]{0,0,1}}%
      \expandafter\def\csname LT3\endcsname{\color[rgb]{1,0,1}}%
      \expandafter\def\csname LT4\endcsname{\color[rgb]{0,1,1}}%
      \expandafter\def\csname LT5\endcsname{\color[rgb]{1,1,0}}%
      \expandafter\def\csname LT6\endcsname{\color[rgb]{0,0,0}}%
      \expandafter\def\csname LT7\endcsname{\color[rgb]{1,0.3,0}}%
      \expandafter\def\csname LT8\endcsname{\color[rgb]{0.5,0.5,0.5}}%
    \else
      \def\colorrgb#1{\color{black}}%
      \def\colorgray#1{\color[gray]{#1}}%
      \expandafter\def\csname LTw\endcsname{\color{white}}%
      \expandafter\def\csname LTb\endcsname{\color{black}}%
      \expandafter\def\csname LTa\endcsname{\color{black}}%
      \expandafter\def\csname LT0\endcsname{\color{black}}%
      \expandafter\def\csname LT1\endcsname{\color{black}}%
      \expandafter\def\csname LT2\endcsname{\color{black}}%
      \expandafter\def\csname LT3\endcsname{\color{black}}%
      \expandafter\def\csname LT4\endcsname{\color{black}}%
      \expandafter\def\csname LT5\endcsname{\color{black}}%
      \expandafter\def\csname LT6\endcsname{\color{black}}%
      \expandafter\def\csname LT7\endcsname{\color{black}}%
      \expandafter\def\csname LT8\endcsname{\color{black}}%
    \fi
  \fi
  \setlength{\unitlength}{0.0500bp}%
  \begin{picture}(11232.00,7342.00)%
    \gplgaddtomacro\gplbacktext{%
      \csname LTb\endcsname%
      \put(991,5953){\makebox(0,0)[r]{\strut{}20}}%
      \put(991,6271){\makebox(0,0)[r]{\strut{}60}}%
      \put(991,6588){\makebox(0,0)[r]{\strut{}100}}%
      \put(991,6906){\makebox(0,0)[r]{\strut{}140}}%
      \put(1997,5495){\makebox(0,0){\strut{}}}%
      \put(3213,5495){\makebox(0,0){\strut{}}}%
      \put(4429,5495){\makebox(0,0){\strut{}}}%
      \put(5645,5495){\makebox(0,0){\strut{}}}%
      \put(6861,5495){\makebox(0,0){\strut{}}}%
      \put(8077,5495){\makebox(0,0){\strut{}}}%
      \put(9293,5495){\makebox(0,0){\strut{}}}%
      \put(10509,5495){\makebox(0,0){\strut{}}}%
      \put(1123,7205){\makebox(0,0){\strut{}2003}}%
      \put(2011,7205){\makebox(0,0){\strut{}2004}}%
      \put(2901,7205){\makebox(0,0){\strut{}2005}}%
      \put(3788,7205){\makebox(0,0){\strut{}2006}}%
      \put(4676,7205){\makebox(0,0){\strut{}2007}}%
      \put(5563,7205){\makebox(0,0){\strut{}2008}}%
      \put(6454,7205){\makebox(0,0){\strut{}2009}}%
      \put(7341,7205){\makebox(0,0){\strut{}2010}}%
      \put(8229,7205){\makebox(0,0){\strut{}2011}}%
      \put(9116,7205){\makebox(0,0){\strut{}2012}}%
      \put(10006,7205){\makebox(0,0){\strut{}2013}}%
      \put(10894,7205){\makebox(0,0){\strut{}2014}}%
      \put(-87,6350){\rotatebox{-270}{\makebox(0,0){\strut{}RXTE-ASM }}}%
      \put(133,6350){\rotatebox{-270}{\makebox(0,0){\strut{} $1.5-12$ keV }}}%
      \put(353,6350){\rotatebox{-270}{\makebox(0,0){\strut{} [cps]}}}%
      \put(6008,7534){\makebox(0,0){\strut{}year}}%
    }%
    \gplgaddtomacro\gplfronttext{%
    }%
    \gplgaddtomacro\gplbacktext{%
      \csname LTb\endcsname%
      \put(991,4502){\makebox(0,0)[r]{\strut{}0.1}}%
      \put(991,5096){\makebox(0,0)[r]{\strut{}1}}%
      \put(991,5690){\makebox(0,0)[r]{\strut{}10}}%
      \put(1997,4224){\makebox(0,0){\strut{}}}%
      \put(3213,4224){\makebox(0,0){\strut{}}}%
      \put(4429,4224){\makebox(0,0){\strut{}}}%
      \put(5645,4224){\makebox(0,0){\strut{}}}%
      \put(6861,4224){\makebox(0,0){\strut{}}}%
      \put(8077,4224){\makebox(0,0){\strut{}}}%
      \put(9293,4224){\makebox(0,0){\strut{}}}%
      \put(10509,4224){\makebox(0,0){\strut{}}}%
      \put(1123,5935){\makebox(0,0){\strut{}}}%
      \put(2011,5935){\makebox(0,0){\strut{}}}%
      \put(2901,5935){\makebox(0,0){\strut{}}}%
      \put(3788,5935){\makebox(0,0){\strut{}}}%
      \put(4676,5935){\makebox(0,0){\strut{}}}%
      \put(5563,5935){\makebox(0,0){\strut{}}}%
      \put(6454,5935){\makebox(0,0){\strut{}}}%
      \put(7341,5935){\makebox(0,0){\strut{}}}%
      \put(8229,5935){\makebox(0,0){\strut{}}}%
      \put(9116,5935){\makebox(0,0){\strut{}}}%
      \put(10006,5935){\makebox(0,0){\strut{}}}%
      \put(10894,5935){\makebox(0,0){\strut{}}}%
      \put(-87,5079){\rotatebox{-270}{\makebox(0,0){\strut{}RXTE-ASM }}}%
      \put(133,5079){\rotatebox{-270}{\makebox(0,0){\strut{} hardness }}}%
      \put(353,5079){\rotatebox{-270}{\makebox(0,0){\strut{} (C/A)}}}%
    }%
    \gplgaddtomacro\gplfronttext{%
    }%
    \gplgaddtomacro\gplbacktext{%
      \csname LTb\endcsname%
      \put(991,3428){\makebox(0,0)[r]{\strut{}1}}%
      \put(991,3682){\makebox(0,0)[r]{\strut{}2}}%
      \put(991,3937){\makebox(0,0)[r]{\strut{}3}}%
      \put(991,4191){\makebox(0,0)[r]{\strut{}4}}%
      \put(1997,2954){\makebox(0,0){\strut{}}}%
      \put(3213,2954){\makebox(0,0){\strut{}}}%
      \put(4429,2954){\makebox(0,0){\strut{}}}%
      \put(5645,2954){\makebox(0,0){\strut{}}}%
      \put(6861,2954){\makebox(0,0){\strut{}}}%
      \put(8077,2954){\makebox(0,0){\strut{}}}%
      \put(9293,2954){\makebox(0,0){\strut{}}}%
      \put(10509,2954){\makebox(0,0){\strut{}}}%
      \put(1123,4665){\makebox(0,0){\strut{}}}%
      \put(2011,4665){\makebox(0,0){\strut{}}}%
      \put(2901,4665){\makebox(0,0){\strut{}}}%
      \put(3788,4665){\makebox(0,0){\strut{}}}%
      \put(4676,4665){\makebox(0,0){\strut{}}}%
      \put(5563,4665){\makebox(0,0){\strut{}}}%
      \put(6454,4665){\makebox(0,0){\strut{}}}%
      \put(7341,4665){\makebox(0,0){\strut{}}}%
      \put(8229,4665){\makebox(0,0){\strut{}}}%
      \put(9116,4665){\makebox(0,0){\strut{}}}%
      \put(10006,4665){\makebox(0,0){\strut{}}}%
      \put(10894,4665){\makebox(0,0){\strut{}}}%
      \put(-87,3809){\rotatebox{-270}{\makebox(0,0){\strut{}MAXI }}}%
      \put(133,3809){\rotatebox{-270}{\makebox(0,0){\strut{} $2-20$ keV }}}%
      \put(353,3809){\rotatebox{-270}{\makebox(0,0){\strut{} [cps/$\mathrm{cm}^2$]}}}%
    }%
    \gplgaddtomacro\gplfronttext{%
    }%
    \gplgaddtomacro\gplbacktext{%
      \csname LTb\endcsname%
      \put(991,2159){\makebox(0,0)[r]{\strut{}0.0}}%
      \put(991,2413){\makebox(0,0)[r]{\strut{}0.1}}%
      \put(991,2668){\makebox(0,0)[r]{\strut{}0.2}}%
      \put(991,2922){\makebox(0,0)[r]{\strut{}0.3}}%
      \put(1997,1685){\makebox(0,0){\strut{}}}%
      \put(3213,1685){\makebox(0,0){\strut{}}}%
      \put(4429,1685){\makebox(0,0){\strut{}}}%
      \put(5645,1685){\makebox(0,0){\strut{}}}%
      \put(6861,1685){\makebox(0,0){\strut{}}}%
      \put(8077,1685){\makebox(0,0){\strut{}}}%
      \put(9293,1685){\makebox(0,0){\strut{}}}%
      \put(10509,1685){\makebox(0,0){\strut{}}}%
      \put(1123,3396){\makebox(0,0){\strut{}}}%
      \put(2011,3396){\makebox(0,0){\strut{}}}%
      \put(2901,3396){\makebox(0,0){\strut{}}}%
      \put(3788,3396){\makebox(0,0){\strut{}}}%
      \put(4676,3396){\makebox(0,0){\strut{}}}%
      \put(5563,3396){\makebox(0,0){\strut{}}}%
      \put(6454,3396){\makebox(0,0){\strut{}}}%
      \put(7341,3396){\makebox(0,0){\strut{}}}%
      \put(8229,3396){\makebox(0,0){\strut{}}}%
      \put(9116,3396){\makebox(0,0){\strut{}}}%
      \put(10006,3396){\makebox(0,0){\strut{}}}%
      \put(10894,3396){\makebox(0,0){\strut{}}}%
      \put(-87,2540){\rotatebox{-270}{\makebox(0,0){\strut{}Swift-BAT }}}%
      \put(133,2540){\rotatebox{-270}{\makebox(0,0){\strut{} $15-50$ keV }}}%
      \put(353,2540){\rotatebox{-270}{\makebox(0,0){\strut{} [cps/$\mathrm{cm}^2$]}}}%
    }%
    \gplgaddtomacro\gplfronttext{%
    }%
    \gplgaddtomacro\gplbacktext{%
      \csname LTb\endcsname%
      \put(1997,415){\makebox(0,0){\strut{}53000}}%
      \put(3213,415){\makebox(0,0){\strut{}53500}}%
      \put(4429,415){\makebox(0,0){\strut{}54000}}%
      \put(5645,415){\makebox(0,0){\strut{}54500}}%
      \put(6861,415){\makebox(0,0){\strut{}55000}}%
      \put(8077,415){\makebox(0,0){\strut{}55500}}%
      \put(9293,415){\makebox(0,0){\strut{}56000}}%
      \put(10509,415){\makebox(0,0){\strut{}56500}}%
      \put(1123,2126){\makebox(0,0){\strut{}}}%
      \put(2011,2126){\makebox(0,0){\strut{}}}%
      \put(2901,2126){\makebox(0,0){\strut{}}}%
      \put(3788,2126){\makebox(0,0){\strut{}}}%
      \put(4676,2126){\makebox(0,0){\strut{}}}%
      \put(5563,2126){\makebox(0,0){\strut{}}}%
      \put(6454,2126){\makebox(0,0){\strut{}}}%
      \put(7341,2126){\makebox(0,0){\strut{}}}%
      \put(8229,2126){\makebox(0,0){\strut{}}}%
      \put(9116,2126){\makebox(0,0){\strut{}}}%
      \put(10006,2126){\makebox(0,0){\strut{}}}%
      \put(10894,2126){\makebox(0,0){\strut{}}}%
      \put(-87,1270){\rotatebox{-270}{\makebox(0,0){\strut{}Observatory}}}%
      \put(6008,85){\makebox(0,0){\strut{}MJD}}%
    }%
    \gplgaddtomacro\gplfronttext{%
      \csname LTb\endcsname%
      \put(-126814,115025){\makebox(0,0)[l]{\strut{}$0.05$}}%
    }%
    \gplbacktext
    \put(0,0){\includegraphics{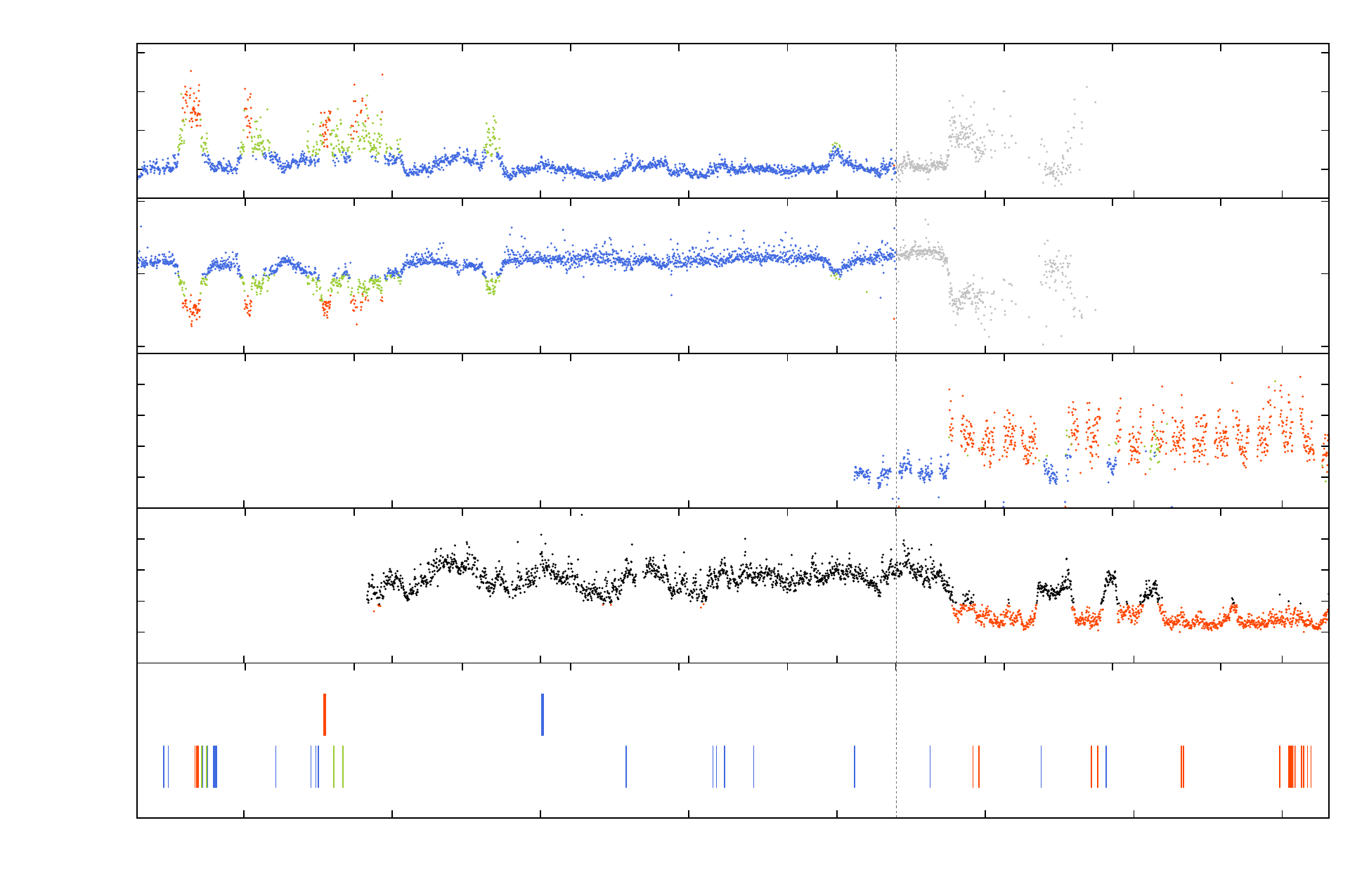}}%
    \gplfronttext
  \end{picture}%
\endgroup

%% file: DTfig007.tex
\begingroup
  \makeatletter
  \providecommand\color[2][]{%
    \GenericError{(gnuplot) \space\space\space\@spaces}{%
      Package color not loaded in conjunction with
      terminal option `colourtext'%
    }{See the gnuplot documentation for explanation.%
    }{Either use 'blacktext' in gnuplot or load the package
      color.sty in LaTeX.}%
    \renewcommand\color[2][]{}%
  }%
  \providecommand\includegraphics[2][]{%
    \GenericError{(gnuplot) \space\space\space\@spaces}{%
      Package graphicx or graphics not loaded%
    }{See the gnuplot documentation for explanation.%
    }{The gnuplot epslatex terminal needs graphicx.sty or graphics.sty.}%
    \renewcommand\includegraphics[2][]{}%
  }%
  \providecommand\rotatebox[2]{#2}%
  \@ifundefined{ifGPcolor}{%
    \newif\ifGPcolor
    \GPcolortrue
  }{}%
  \@ifundefined{ifGPblacktext}{%
    \newif\ifGPblacktext
    \GPblacktextfalse
  }{}%
  \let\gplgaddtomacro\g@addto@macro
  \gdef\gplbacktext{}%
  \gdef\gplfronttext{}%
  \makeatother
  \ifGPblacktext
    \def\colorrgb#1{}%
    \def\colorgray#1{}%
  \else
    \ifGPcolor
      \def\colorrgb#1{\color[rgb]{#1}}%
      \def\colorgray#1{\color[gray]{#1}}%
      \expandafter\def\csname LTw\endcsname{\color{white}}%
      \expandafter\def\csname LTb\endcsname{\color{black}}%
      \expandafter\def\csname LTa\endcsname{\color{black}}%
      \expandafter\def\csname LT0\endcsname{\color[rgb]{1,0,0}}%
      \expandafter\def\csname LT1\endcsname{\color[rgb]{0,1,0}}%
      \expandafter\def\csname LT2\endcsname{\color[rgb]{0,0,1}}%
      \expandafter\def\csname LT3\endcsname{\color[rgb]{1,0,1}}%
      \expandafter\def\csname LT4\endcsname{\color[rgb]{0,1,1}}%
      \expandafter\def\csname LT5\endcsname{\color[rgb]{1,1,0}}%
      \expandafter\def\csname LT6\endcsname{\color[rgb]{0,0,0}}%
      \expandafter\def\csname LT7\endcsname{\color[rgb]{1,0.3,0}}%
      \expandafter\def\csname LT8\endcsname{\color[rgb]{0.5,0.5,0.5}}%
    \else
      \def\colorrgb#1{\color{black}}%
      \def\colorgray#1{\color[gray]{#1}}%
      \expandafter\def\csname LTw\endcsname{\color{white}}%
      \expandafter\def\csname LTb\endcsname{\color{black}}%
      \expandafter\def\csname LTa\endcsname{\color{black}}%
      \expandafter\def\csname LT0\endcsname{\color{black}}%
      \expandafter\def\csname LT1\endcsname{\color{black}}%
      \expandafter\def\csname LT2\endcsname{\color{black}}%
      \expandafter\def\csname LT3\endcsname{\color{black}}%
      \expandafter\def\csname LT4\endcsname{\color{black}}%
      \expandafter\def\csname LT5\endcsname{\color{black}}%
      \expandafter\def\csname LT6\endcsname{\color{black}}%
      \expandafter\def\csname LT7\endcsname{\color{black}}%
      \expandafter\def\csname LT8\endcsname{\color{black}}%
    \fi
  \fi
  \setlength{\unitlength}{0.0500bp}%
  \begin{picture}(11232.00,4664.00)%
    \gplgaddtomacro\gplbacktext{%
      \csname LTb\endcsname%
      \put(280,3147){\makebox(0,0){\strut{}}}%
      \put(1347,3147){\makebox(0,0){\strut{}}}%
      \put(2414,3147){\makebox(0,0){\strut{}}}%
      \put(3481,3147){\makebox(0,0){\strut{}}}%
      \put(4548,3147){\makebox(0,0){\strut{}}}%
      \put(5615,3147){\makebox(0,0){\strut{}}}%
      \put(6682,3147){\makebox(0,0){\strut{}}}%
      \put(7749,3147){\makebox(0,0){\strut{}}}%
      \put(8816,3147){\makebox(0,0){\strut{}}}%
      \put(9883,3147){\makebox(0,0){\strut{}}}%
      \put(10950,3147){\makebox(0,0){\strut{}}}%
      \put(5615,4458){\makebox(0,0){\strut{}Low/Hard state}}%
    }%
    \gplgaddtomacro\gplfronttext{%
    }%
    \gplgaddtomacro\gplbacktext{%
      \csname LTb\endcsname%
      \put(280,1748){\makebox(0,0){\strut{}}}%
      \put(1347,1748){\makebox(0,0){\strut{}}}%
      \put(2414,1748){\makebox(0,0){\strut{}}}%
      \put(3481,1748){\makebox(0,0){\strut{}}}%
      \put(4548,1748){\makebox(0,0){\strut{}}}%
      \put(5615,1748){\makebox(0,0){\strut{}}}%
      \put(6682,1748){\makebox(0,0){\strut{}}}%
      \put(7749,1748){\makebox(0,0){\strut{}}}%
      \put(8816,1748){\makebox(0,0){\strut{}}}%
      \put(9883,1748){\makebox(0,0){\strut{}}}%
      \put(10950,1748){\makebox(0,0){\strut{}}}%
      \put(5615,3059){\makebox(0,0){\strut{}Intermediate state}}%
    }%
    \gplgaddtomacro\gplfronttext{%
    }%
    \gplgaddtomacro\gplbacktext{%
      \csname LTb\endcsname%
      \put(280,349){\makebox(0,0){\strut{}0}}%
      \put(1347,349){\makebox(0,0){\strut{}0.1}}%
      \put(2414,349){\makebox(0,0){\strut{}0.2}}%
      \put(3481,349){\makebox(0,0){\strut{}0.3}}%
      \put(4548,349){\makebox(0,0){\strut{}0.4}}%
      \put(5615,349){\makebox(0,0){\strut{}0.5}}%
      \put(6682,349){\makebox(0,0){\strut{}0.6}}%
      \put(7749,349){\makebox(0,0){\strut{}0.7}}%
      \put(8816,349){\makebox(0,0){\strut{}0.8}}%
      \put(9883,349){\makebox(0,0){\strut{}0.9}}%
      \put(10950,349){\makebox(0,0){\strut{}1}}%
      \put(5615,19){\makebox(0,0){\strut{}$\Phi$}}%
      \put(5615,1661){\makebox(0,0){\strut{}High/Soft state}}%
    }%
    \gplgaddtomacro\gplfronttext{%
    }%
    \gplbacktext
    \put(0,0){\includegraphics{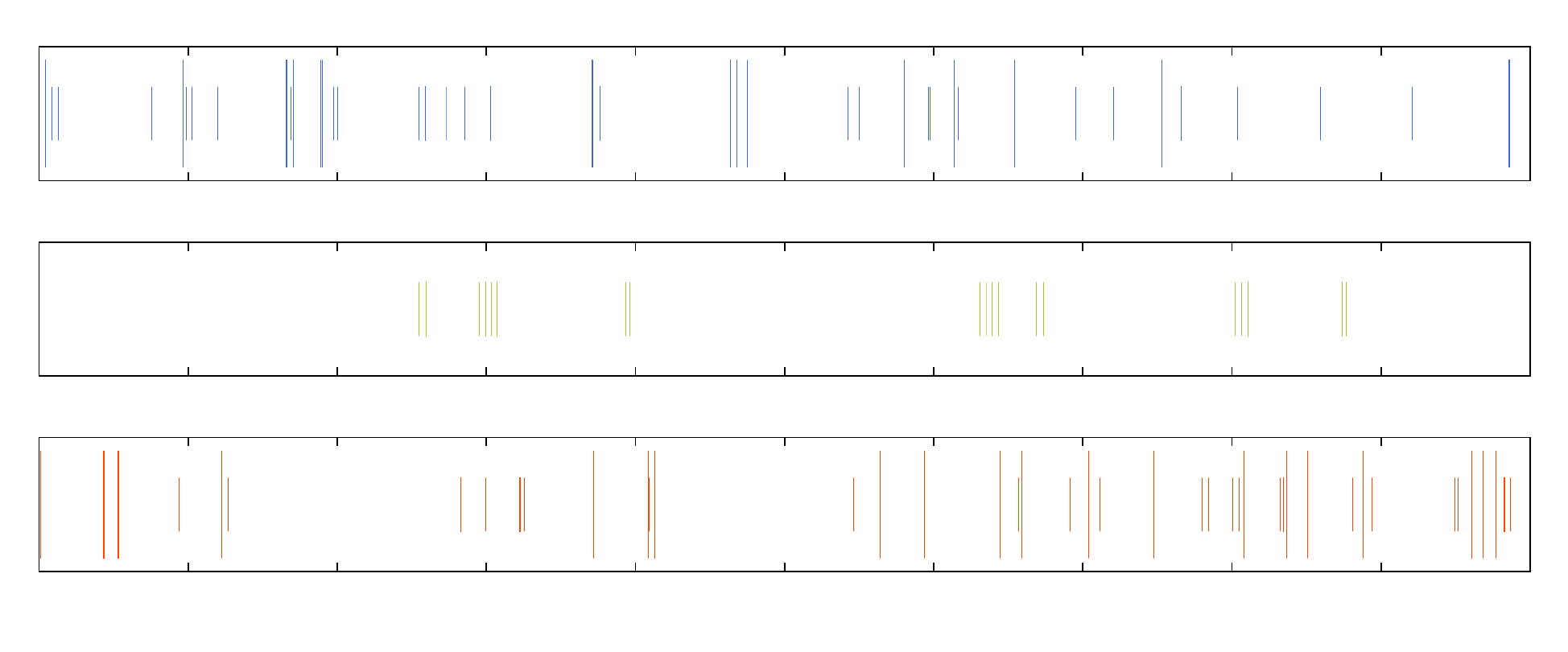}}%
    \gplfronttext
  \end{picture}%
\endgroup

%% file: DTfig008.tex
\begingroup
  \makeatletter
  \providecommand\color[2][]{%
    \GenericError{(gnuplot) \space\space\space\@spaces}{%
      Package color not loaded in conjunction with
      terminal option `colourtext'%
    }{See the gnuplot documentation for explanation.%
    }{Either use 'blacktext' in gnuplot or load the package
      color.sty in LaTeX.}%
    \renewcommand\color[2][]{}%
  }%
  \providecommand\includegraphics[2][]{%
    \GenericError{(gnuplot) \space\space\space\@spaces}{%
      Package graphicx or graphics not loaded%
    }{See the gnuplot documentation for explanation.%
    }{The gnuplot epslatex terminal needs graphicx.sty or graphics.sty.}%
    \renewcommand\includegraphics[2][]{}%
  }%
  \providecommand\rotatebox[2]{#2}%
  \@ifundefined{ifGPcolor}{%
    \newif\ifGPcolor
    \GPcolortrue
  }{}%
  \@ifundefined{ifGPblacktext}{%
    \newif\ifGPblacktext
    \GPblacktextfalse
  }{}%
  \let\gplgaddtomacro\g@addto@macro
  \gdef\gplbacktext{}%
  \gdef\gplfronttext{}%
  \makeatother
  \ifGPblacktext
    \def\colorrgb#1{}%
    \def\colorgray#1{}%
  \else
    \ifGPcolor
      \def\colorrgb#1{\color[rgb]{#1}}%
      \def\colorgray#1{\color[gray]{#1}}%
      \expandafter\def\csname LTw\endcsname{\color{white}}%
      \expandafter\def\csname LTb\endcsname{\color{black}}%
      \expandafter\def\csname LTa\endcsname{\color{black}}%
      \expandafter\def\csname LT0\endcsname{\color[rgb]{1,0,0}}%
      \expandafter\def\csname LT1\endcsname{\color[rgb]{0,1,0}}%
      \expandafter\def\csname LT2\endcsname{\color[rgb]{0,0,1}}%
      \expandafter\def\csname LT3\endcsname{\color[rgb]{1,0,1}}%
      \expandafter\def\csname LT4\endcsname{\color[rgb]{0,1,1}}%
      \expandafter\def\csname LT5\endcsname{\color[rgb]{1,1,0}}%
      \expandafter\def\csname LT6\endcsname{\color[rgb]{0,0,0}}%
      \expandafter\def\csname LT7\endcsname{\color[rgb]{1,0.3,0}}%
      \expandafter\def\csname LT8\endcsname{\color[rgb]{0.5,0.5,0.5}}%
    \else
      \def\colorrgb#1{\color{black}}%
      \def\colorgray#1{\color[gray]{#1}}%
      \expandafter\def\csname LTw\endcsname{\color{white}}%
      \expandafter\def\csname LTb\endcsname{\color{black}}%
      \expandafter\def\csname LTa\endcsname{\color{black}}%
      \expandafter\def\csname LT0\endcsname{\color{black}}%
      \expandafter\def\csname LT1\endcsname{\color{black}}%
      \expandafter\def\csname LT2\endcsname{\color{black}}%
      \expandafter\def\csname LT3\endcsname{\color{black}}%
      \expandafter\def\csname LT4\endcsname{\color{black}}%
      \expandafter\def\csname LT5\endcsname{\color{black}}%
      \expandafter\def\csname LT6\endcsname{\color{black}}%
      \expandafter\def\csname LT7\endcsname{\color{black}}%
      \expandafter\def\csname LT8\endcsname{\color{black}}%
    \fi
  \fi
  \setlength{\unitlength}{0.0500bp}%
  \begin{picture}(11232.00,8640.00)%
    \gplgaddtomacro\gplbacktext{%
      \csname LTb\endcsname%
      \put(943,471){\makebox(0,0){\strut{}6350}}%
      \put(2100,471){\makebox(0,0){\strut{}6450}}%
      \put(3257,471){\makebox(0,0){\strut{}6550}}%
      \put(4413,471){\makebox(0,0){\strut{}6650}}%
      \put(5570,471){\makebox(0,0){\strut{}6750}}%
      \put(723,4578){\rotatebox{-270}{\makebox(0,0){\strut{}normalized flux}}}%
      \put(3256,141){\makebox(0,0){\strut{}$\lambda(\mathrm{\mbox{\r{A}}})$}}%
      \put(5223,7945){\makebox(0,0){\tiny{0.986}}}%
      \put(5223,7385){\makebox(0,0){\tiny{0.757}}}%
      \put(5223,6824){\makebox(0,0){\tiny{0.654}}}%
      \put(5223,6262){\makebox(0,0){\tiny{0.614}}}%
      \put(5223,5701){\makebox(0,0){\tiny{0.580}}}%
      \put(5223,5140){\makebox(0,0){\tiny{0.475}}}%
      \put(5223,4579){\makebox(0,0){\tiny{0.468}}}%
      \put(5223,4017){\makebox(0,0){\tiny{0.463}}}%
      \put(5223,3456){\makebox(0,0){\tiny{0.374}}}%
      \put(5223,2895){\makebox(0,0){\tiny{0.194}}}%
      \put(5223,2333){\makebox(0,0){\tiny{0.170}}}%
      \put(5223,1772){\makebox(0,0){\tiny{0.097}}}%
      \put(5223,1212){\makebox(0,0){\tiny{0.004}}}%
      \put(1521,7945){\makebox(0,0){\tiny{2003-08-10}}}%
      \put(1521,7385){\makebox(0,0){\tiny{2003-09-16}}}%
      \put(1521,6824){\makebox(0,0){\tiny{2003-09-21}}}%
      \put(1521,6262){\makebox(0,0){\tiny{2003-04-01}}}%
      \put(1521,5701){\makebox(0,0){\tiny{2003-09-15}}}%
      \put(1521,5140){\makebox(0,0){\tiny{2003-09-20b}}}%
      \put(1521,4579){\makebox(0,0){\tiny{2003-09-20a}}}%
      \put(1521,4017){\makebox(0,0){\tiny{2003-04-17}}}%
      \put(1521,3456){\makebox(0,0){\tiny{2003-09-25}}}%
      \put(1521,2895){\makebox(0,0){\tiny{2003-09-24}}}%
      \put(1521,2333){\makebox(0,0){\tiny{2003-08-11}}}%
      \put(1521,1772){\makebox(0,0){\tiny{2003-09-18}}}%
      \put(1521,1212){\makebox(0,0){\tiny{2003-08-26}}}%
    }%
    \gplgaddtomacro\gplfronttext{%
    }%
    \gplgaddtomacro\gplbacktext{%
      \csname LTb\endcsname%
      \put(943,471){\makebox(0,0){\strut{}6350}}%
      \put(2100,471){\makebox(0,0){\strut{}6450}}%
      \put(3257,471){\makebox(0,0){\strut{}6550}}%
      \put(4413,471){\makebox(0,0){\strut{}6650}}%
      \put(5570,471){\makebox(0,0){\strut{}6750}}%
      \put(723,4578){\rotatebox{-270}{\makebox(0,0){\strut{}normalized flux}}}%
      \put(3256,141){\makebox(0,0){\strut{}$\lambda(\mathrm{\mbox{\r{A}}})$}}%
      \put(5223,7945){\makebox(0,0){\tiny{0.986}}}%
      \put(5223,7385){\makebox(0,0){\tiny{0.757}}}%
      \put(5223,6824){\makebox(0,0){\tiny{0.654}}}%
      \put(5223,6262){\makebox(0,0){\tiny{0.614}}}%
      \put(5223,5701){\makebox(0,0){\tiny{0.580}}}%
      \put(5223,5140){\makebox(0,0){\tiny{0.475}}}%
      \put(5223,4579){\makebox(0,0){\tiny{0.468}}}%
      \put(5223,4017){\makebox(0,0){\tiny{0.463}}}%
      \put(5223,3456){\makebox(0,0){\tiny{0.374}}}%
      \put(5223,2895){\makebox(0,0){\tiny{0.194}}}%
      \put(5223,2333){\makebox(0,0){\tiny{0.170}}}%
      \put(5223,1772){\makebox(0,0){\tiny{0.097}}}%
      \put(5223,1212){\makebox(0,0){\tiny{0.004}}}%
      \put(1521,7945){\makebox(0,0){\tiny{2003-08-10}}}%
      \put(1521,7385){\makebox(0,0){\tiny{2003-09-16}}}%
      \put(1521,6824){\makebox(0,0){\tiny{2003-09-21}}}%
      \put(1521,6262){\makebox(0,0){\tiny{2003-04-01}}}%
      \put(1521,5701){\makebox(0,0){\tiny{2003-09-15}}}%
      \put(1521,5140){\makebox(0,0){\tiny{2003-09-20b}}}%
      \put(1521,4579){\makebox(0,0){\tiny{2003-09-20a}}}%
      \put(1521,4017){\makebox(0,0){\tiny{2003-04-17}}}%
      \put(1521,3456){\makebox(0,0){\tiny{2003-09-25}}}%
      \put(1521,2895){\makebox(0,0){\tiny{2003-09-24}}}%
      \put(1521,2333){\makebox(0,0){\tiny{2003-08-11}}}%
      \put(1521,1772){\makebox(0,0){\tiny{2003-09-18}}}%
      \put(1521,1212){\makebox(0,0){\tiny{2003-08-26}}}%
      \put(2687,8276){\makebox(0,0){\scriptsize{telluric lines}}}%
      \put(3404,8276){\makebox(0,0){\scriptsize{$\mathrm{H}\alpha$}}}%
      \put(4737,8276){\makebox(0,0){\scriptsize{He \rom{1}}}}%
      \put(3997,8276){\makebox(0,0){\scriptsize{DIB}}}%
    }%
    \gplgaddtomacro\gplfronttext{%
    }%
    \gplgaddtomacro\gplbacktext{%
      \csname LTb\endcsname%
      \put(6446,471){\makebox(0,0){\strut{}6350}}%
      \put(7768,471){\makebox(0,0){\strut{}6450}}%
      \put(9091,471){\makebox(0,0){\strut{}6550}}%
      \put(10413,471){\makebox(0,0){\strut{}6650}}%
      \put(6226,4578){\rotatebox{-270}{\makebox(0,0){\strut{}normalized flux}}}%
      \put(8760,141){\makebox(0,0){\strut{}$\lambda(\mathrm{\mbox{\r{A}}})$}}%
      \put(10281,7901){\makebox(0,0){\tiny{0.977}}}%
      \put(10281,7547){\makebox(0,0){\tiny{0.968}}}%
      \put(10281,7194){\makebox(0,0){\tiny{0.960}}}%
      \put(10281,6840){\makebox(0,0){\tiny{0.887}}}%
      \put(10281,6487){\makebox(0,0){\tiny{0.850}}}%
      \put(10281,6134){\makebox(0,0){\tiny{0.836}}}%
      \put(10281,5780){\makebox(0,0){\tiny{0.807}}}%
      \put(10281,5427){\makebox(0,0){\tiny{0.747}}}%
      \put(10281,5073){\makebox(0,0){\tiny{0.703}}}%
      \put(10281,4720){\makebox(0,0){\tiny{0.659}}}%
      \put(10281,4366){\makebox(0,0){\tiny{0.644}}}%
      \put(10281,4013){\makebox(0,0){\tiny{0.593}}}%
      \put(10281,3660){\makebox(0,0){\tiny{0.563}}}%
      \put(10281,3306){\makebox(0,0){\tiny{0.413}}}%
      \put(10281,2953){\makebox(0,0){\tiny{0.408}}}%
      \put(10281,2599){\makebox(0,0){\tiny{0.371}}}%
      \put(10281,2246){\makebox(0,0){\tiny{0.122}}}%
      \put(10281,1893){\makebox(0,0){\tiny{0.053}}}%
      \put(10281,1539){\makebox(0,0){\tiny{0.043}}}%
      \put(10281,1186){\makebox(0,0){\tiny{0.001}}}%
      \put(7107,7901){\makebox(0,0){\tiny{2013-10-19}}}%
      \put(7107,7547){\makebox(0,0){\tiny{2013-08-24}}}%
      \put(7107,7194){\makebox(0,0){\tiny{2013-04-09}}}%
      \put(7107,6840){\makebox(0,0){\tiny{2013-08-29}}}%
      \put(7107,6487){\makebox(0,0){\tiny{2013-10-07}}}%
      \put(7107,6134){\makebox(0,0){\tiny{2013-09-09}}}%
      \put(7107,5780){\makebox(0,0){\tiny{2013-08-23}}}%
      \put(7107,5427){\makebox(0,0){\tiny{2013-08-17}}}%
      \put(7107,5073){\makebox(0,0){\tiny{2013-08-28}}}%
      \put(7107,4720){\makebox(0,0){\tiny{2013-10-06}}}%
      \put(7107,4366){\makebox(0,0){\tiny{2013-08-22}}}%
      \put(7107,4013){\makebox(0,0){\tiny{2013-09-30}}}%
      \put(7107,3660){\makebox(0,0){\tiny{2013-08-16}}}%
      \put(7107,3306){\makebox(0,0){\tiny{2013-09-29}}}%
      \put(7107,2953){\makebox(0,0){\tiny{2013-07-19}}}%
      \put(7107,2599){\makebox(0,0){\tiny{2013-07-18}}}%
      \put(7107,2246){\makebox(0,0){\tiny{2013-10-31}}}%
      \put(7107,1893){\makebox(0,0){\tiny{2013-09-27}}}%
      \put(7107,1539){\makebox(0,0){\tiny{2013-07-17}}}%
      \put(7107,1186){\makebox(0,0){\tiny{2013-08-10}}}%
    }%
    \gplgaddtomacro\gplfronttext{%
    }%
    \gplgaddtomacro\gplbacktext{%
      \csname LTb\endcsname%
      \put(6446,471){\makebox(0,0){\strut{}6350}}%
      \put(7768,471){\makebox(0,0){\strut{}6450}}%
      \put(9091,471){\makebox(0,0){\strut{}6550}}%
      \put(10413,471){\makebox(0,0){\strut{}6650}}%
      \put(6226,4578){\rotatebox{-270}{\makebox(0,0){\strut{}normalized flux}}}%
      \put(8760,141){\makebox(0,0){\strut{}$\lambda(\mathrm{\mbox{\r{A}}})$}}%
      \put(10281,7901){\makebox(0,0){\tiny{0.977}}}%
      \put(10281,7547){\makebox(0,0){\tiny{0.968}}}%
      \put(10281,7194){\makebox(0,0){\tiny{0.960}}}%
      \put(10281,6840){\makebox(0,0){\tiny{0.887}}}%
      \put(10281,6487){\makebox(0,0){\tiny{0.850}}}%
      \put(10281,6134){\makebox(0,0){\tiny{0.836}}}%
      \put(10281,5780){\makebox(0,0){\tiny{0.807}}}%
      \put(10281,5427){\makebox(0,0){\tiny{0.747}}}%
      \put(10281,5073){\makebox(0,0){\tiny{0.703}}}%
      \put(10281,4720){\makebox(0,0){\tiny{0.659}}}%
      \put(10281,4366){\makebox(0,0){\tiny{0.644}}}%
      \put(10281,4013){\makebox(0,0){\tiny{0.593}}}%
      \put(10281,3660){\makebox(0,0){\tiny{0.563}}}%
      \put(10281,3306){\makebox(0,0){\tiny{0.413}}}%
      \put(10281,2953){\makebox(0,0){\tiny{0.408}}}%
      \put(10281,2599){\makebox(0,0){\tiny{0.371}}}%
      \put(10281,2246){\makebox(0,0){\tiny{0.122}}}%
      \put(10281,1893){\makebox(0,0){\tiny{0.053}}}%
      \put(10281,1539){\makebox(0,0){\tiny{0.043}}}%
      \put(10281,1186){\makebox(0,0){\tiny{0.001}}}%
      \put(7107,7901){\makebox(0,0){\tiny{2013-10-19}}}%
      \put(7107,7547){\makebox(0,0){\tiny{2013-08-24}}}%
      \put(7107,7194){\makebox(0,0){\tiny{2013-04-09}}}%
      \put(7107,6840){\makebox(0,0){\tiny{2013-08-29}}}%
      \put(7107,6487){\makebox(0,0){\tiny{2013-10-07}}}%
      \put(7107,6134){\makebox(0,0){\tiny{2013-09-09}}}%
      \put(7107,5780){\makebox(0,0){\tiny{2013-08-23}}}%
      \put(7107,5427){\makebox(0,0){\tiny{2013-08-17}}}%
      \put(7107,5073){\makebox(0,0){\tiny{2013-08-28}}}%
      \put(7107,4720){\makebox(0,0){\tiny{2013-10-06}}}%
      \put(7107,4366){\makebox(0,0){\tiny{2013-08-22}}}%
      \put(7107,4013){\makebox(0,0){\tiny{2013-09-30}}}%
      \put(7107,3660){\makebox(0,0){\tiny{2013-08-16}}}%
      \put(7107,3306){\makebox(0,0){\tiny{2013-09-29}}}%
      \put(7107,2953){\makebox(0,0){\tiny{2013-07-19}}}%
      \put(7107,2599){\makebox(0,0){\tiny{2013-07-18}}}%
      \put(7107,2246){\makebox(0,0){\tiny{2013-10-31}}}%
      \put(7107,1893){\makebox(0,0){\tiny{2013-09-27}}}%
      \put(7107,1539){\makebox(0,0){\tiny{2013-07-17}}}%
      \put(7107,1186){\makebox(0,0){\tiny{2013-08-10}}}%
      \put(8439,8282){\makebox(0,0){\scriptsize{telluric lines}}}%
      \put(9259,8282){\makebox(0,0){\scriptsize{$\mathrm{H}\alpha$}}}%
      \put(10783,8282){\makebox(0,0){\scriptsize{He \rom{1}}}}%
      \put(9937,8282){\makebox(0,0){\scriptsize{DIB}}}%
    }%
    \gplgaddtomacro\gplfronttext{%
    }%
    \gplbacktext
    \put(0,0){\includegraphics{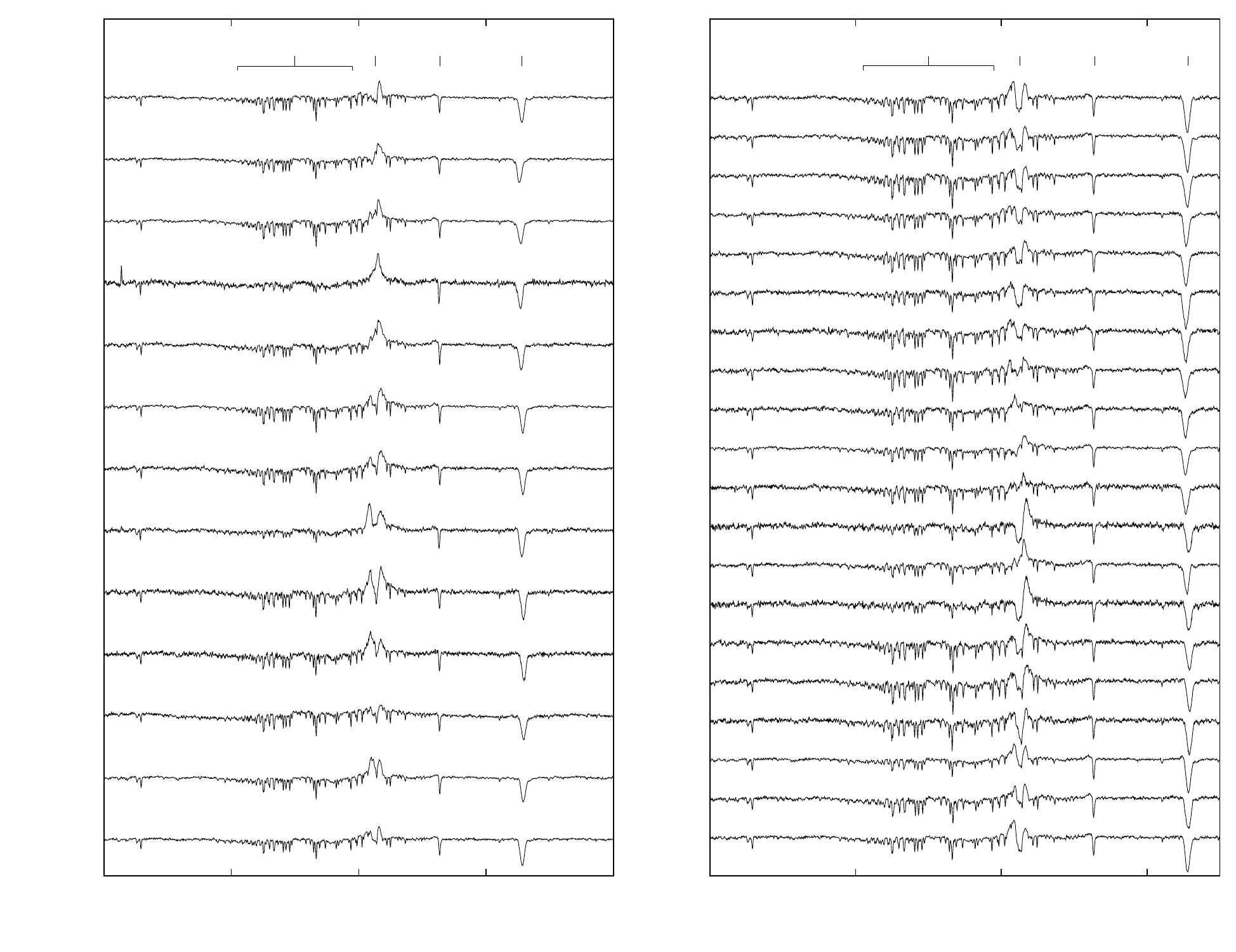}}%
    \gplfronttext
  \end{picture}%
\endgroup

%% file: TP_11.tex
\begingroup
  \makeatletter
  \providecommand\color[2][]{%
    \GenericError{(gnuplot) \space\space\space\@spaces}{%
      Package color not loaded in conjunction with
      terminal option `colourtext'%
    }{See the gnuplot documentation for explanation.%
    }{Either use 'blacktext' in gnuplot or load the package
      color.sty in LaTeX.}%
    \renewcommand\color[2][]{}%
  }%
  \providecommand\includegraphics[2][]{%
    \GenericError{(gnuplot) \space\space\space\@spaces}{%
      Package graphicx or graphics not loaded%
    }{See the gnuplot documentation for explanation.%
    }{The gnuplot epslatex terminal needs graphicx.sty or graphics.sty.}%
    \renewcommand\includegraphics[2][]{}%
  }%
  \providecommand\rotatebox[2]{#2}%
  \@ifundefined{ifGPcolor}{%
    \newif\ifGPcolor
    \GPcolortrue
  }{}%
  \@ifundefined{ifGPblacktext}{%
    \newif\ifGPblacktext
    \GPblacktextfalse
  }{}%
  \let\gplgaddtomacro\g@addto@macro
  \gdef\gplbacktext{}%
  \gdef\gplfronttext{}%
  \makeatother
  \ifGPblacktext
    \def\colorrgb#1{}%
    \def\colorgray#1{}%
  \else
    \ifGPcolor
      \def\colorrgb#1{\color[rgb]{#1}}%
      \def\colorgray#1{\color[gray]{#1}}%
      \expandafter\def\csname LTw\endcsname{\color{white}}%
      \expandafter\def\csname LTb\endcsname{\color{black}}%
      \expandafter\def\csname LTa\endcsname{\color{black}}%
      \expandafter\def\csname LT0\endcsname{\color[rgb]{1,0,0}}%
      \expandafter\def\csname LT1\endcsname{\color[rgb]{0,1,0}}%
      \expandafter\def\csname LT2\endcsname{\color[rgb]{0,0,1}}%
      \expandafter\def\csname LT3\endcsname{\color[rgb]{1,0,1}}%
      \expandafter\def\csname LT4\endcsname{\color[rgb]{0,1,1}}%
      \expandafter\def\csname LT5\endcsname{\color[rgb]{1,1,0}}%
      \expandafter\def\csname LT6\endcsname{\color[rgb]{0,0,0}}%
      \expandafter\def\csname LT7\endcsname{\color[rgb]{1,0.3,0}}%
      \expandafter\def\csname LT8\endcsname{\color[rgb]{0.5,0.5,0.5}}%
    \else
      \def\colorrgb#1{\color{black}}%
      \def\colorgray#1{\color[gray]{#1}}%
      \expandafter\def\csname LTw\endcsname{\color{white}}%
      \expandafter\def\csname LTb\endcsname{\color{black}}%
      \expandafter\def\csname LTa\endcsname{\color{black}}%
      \expandafter\def\csname LT0\endcsname{\color{black}}%
      \expandafter\def\csname LT1\endcsname{\color{black}}%
      \expandafter\def\csname LT2\endcsname{\color{black}}%
      \expandafter\def\csname LT3\endcsname{\color{black}}%
      \expandafter\def\csname LT4\endcsname{\color{black}}%
      \expandafter\def\csname LT5\endcsname{\color{black}}%
      \expandafter\def\csname LT6\endcsname{\color{black}}%
      \expandafter\def\csname LT7\endcsname{\color{black}}%
      \expandafter\def\csname LT8\endcsname{\color{black}}%
    \fi
  \fi
  \setlength{\unitlength}{0.0500bp}%
  \begin{picture}(11232.00,7342.00)%
    \gplgaddtomacro\gplbacktext{%
    }%
    \gplgaddtomacro\gplfronttext{%
      \csname LTb\endcsname%
      \put(477,3838){\makebox(0,0){\strut{}}}%
      \put(1122,3838){\makebox(0,0){\strut{}}}%
      \put(1777,3838){\makebox(0,0){\strut{}}}%
      \put(2430,3838){\makebox(0,0){\strut{}}}%
      \put(3085,3838){\makebox(0,0){\strut{}}}%
      \put(3739,3838){\makebox(0,0){\strut{}}}%
      \put(305,4124){\makebox(0,0)[r]{\strut{}}}%
      \put(305,4770){\makebox(0,0)[r]{\strut{}-0.5}}%
      \put(305,5425){\makebox(0,0)[r]{\strut{}0.0}}%
      \put(305,6079){\makebox(0,0)[r]{\strut{}0.5}}%
      \put(305,6734){\makebox(0,0)[r]{\strut{}1.0}}%
      \put(1916,6918){\makebox(0,0)[l]{\strut{}$\rho\ [\mathrm{kg}\ \mathrm{m}^{-3}]$}}%
    }%
    \gplgaddtomacro\gplbacktext{%
    }%
    \gplgaddtomacro\gplfronttext{%
      \csname LTb\endcsname%
      \put(3987,3838){\makebox(0,0){\strut{}}}%
      \put(4632,3838){\makebox(0,0){\strut{}}}%
      \put(5287,3838){\makebox(0,0){\strut{}}}%
      \put(5940,3838){\makebox(0,0){\strut{}}}%
      \put(6595,3838){\makebox(0,0){\strut{}}}%
      \put(7249,3838){\makebox(0,0){\strut{}}}%
      \put(3815,4124){\makebox(0,0)[r]{\strut{}}}%
      \put(3815,4770){\makebox(0,0)[r]{\strut{}}}%
      \put(3815,5425){\makebox(0,0)[r]{\strut{}}}%
      \put(3815,6079){\makebox(0,0)[r]{\strut{}}}%
      \put(3815,6734){\makebox(0,0)[r]{\strut{}}}%
      \put(5557,6918){\makebox(0,0)[l]{\strut{}$T\ [\mathrm{K}]$}}%
    }%
    \gplgaddtomacro\gplbacktext{%
    }%
    \gplgaddtomacro\gplfronttext{%
      \csname LTb\endcsname%
      \put(7497,3838){\makebox(0,0){\strut{}}}%
      \put(8142,3838){\makebox(0,0){\strut{}}}%
      \put(8797,3838){\makebox(0,0){\strut{}}}%
      \put(9450,3838){\makebox(0,0){\strut{}}}%
      \put(10105,3838){\makebox(0,0){\strut{}}}%
      \put(10759,3838){\makebox(0,0){\strut{}}}%
      \put(7325,4124){\makebox(0,0)[r]{\strut{}}}%
      \put(7325,4770){\makebox(0,0)[r]{\strut{}}}%
      \put(7325,5425){\makebox(0,0)[r]{\strut{}}}%
      \put(7325,6079){\makebox(0,0)[r]{\strut{}}}%
      \put(7325,6734){\makebox(0,0)[r]{\strut{}}}%
      \put(9111,6918){\makebox(0,0)[l]{\strut{}$n_{\mathrm{H}}$}}%
    }%
    \gplgaddtomacro\gplbacktext{%
    }%
    \gplgaddtomacro\gplfronttext{%
      \csname LTb\endcsname%
      \put(477,1202){\makebox(0,0){\strut{}-1.0}}%
      \put(1122,1202){\makebox(0,0){\strut{}-0.5}}%
      \put(1777,1202){\makebox(0,0){\strut{}0.0}}%
      \put(2430,1202){\makebox(0,0){\strut{}0.5}}%
      \put(3085,1202){\makebox(0,0){\strut{}1.0}}%
      \put(3739,1202){\makebox(0,0){\strut{}1.5}}%
      \put(305,1488){\makebox(0,0)[r]{\strut{}-1.0}}%
      \put(305,2134){\makebox(0,0)[r]{\strut{}-0.5}}%
      \put(305,2789){\makebox(0,0)[r]{\strut{}0.0}}%
      \put(305,3443){\makebox(0,0)[r]{\strut{}0.5}}%
      \put(305,4098){\makebox(0,0)[r]{\strut{}1.0}}%
      \put(472,514){\makebox(0,0){\strut{}}}%
      \put(1632,514){\makebox(0,0){\strut{}$2\times 10^{-9}$}}%
      \put(2793,514){\makebox(0,0){\strut{}$4\times 10^{-9}$}}%
      \put(3954,514){\makebox(0,0){\strut{}$6\times 10^{-9}$}}%
    }%
    \gplgaddtomacro\gplbacktext{%
    }%
    \gplgaddtomacro\gplfronttext{%
      \csname LTb\endcsname%
      \put(3987,1202){\makebox(0,0){\strut{}-1.0}}%
      \put(4632,1202){\makebox(0,0){\strut{}-0.5}}%
      \put(5287,1202){\makebox(0,0){\strut{}0.0}}%
      \put(5940,1202){\makebox(0,0){\strut{}0.5}}%
      \put(6595,1202){\makebox(0,0){\strut{}1.0}}%
      \put(7249,1202){\makebox(0,0){\strut{}1.5}}%
      \put(3815,1488){\makebox(0,0)[r]{\strut{}}}%
      \put(3815,2134){\makebox(0,0)[r]{\strut{}}}%
      \put(3815,2789){\makebox(0,0)[r]{\strut{}}}%
      \put(3815,3443){\makebox(0,0)[r]{\strut{}}}%
      \put(3815,4098){\makebox(0,0)[r]{\strut{}}}%
      \put(4857,514){\makebox(0,0){\strut{}$10^{5}$}}%
      \put(5728,514){\makebox(0,0){\strut{}$10^{6}$}}%
      \put(6598,514){\makebox(0,0){\strut{}$10^{7}$}}%
      \put(7469,514){\makebox(0,0){\strut{}$10^{8}$}}%
    }%
    \gplgaddtomacro\gplbacktext{%
    }%
    \gplgaddtomacro\gplfronttext{%
      \csname LTb\endcsname%
      \put(7497,1202){\makebox(0,0){\strut{}-1.0}}%
      \put(8142,1202){\makebox(0,0){\strut{}-0.5}}%
      \put(8797,1202){\makebox(0,0){\strut{}0.0}}%
      \put(9450,1202){\makebox(0,0){\strut{}0.5}}%
      \put(10105,1202){\makebox(0,0){\strut{}1.0}}%
      \put(10759,1202){\makebox(0,0){\strut{}1.5}}%
      \put(7325,1488){\makebox(0,0)[r]{\strut{}}}%
      \put(7325,2134){\makebox(0,0)[r]{\strut{}}}%
      \put(7325,2789){\makebox(0,0)[r]{\strut{}}}%
      \put(7325,3443){\makebox(0,0)[r]{\strut{}}}%
      \put(7325,4098){\makebox(0,0)[r]{\strut{}}}%
      \put(7994,514){\makebox(0,0){\strut{}0.002}}%
      \put(8989,514){\makebox(0,0){\strut{}0.006}}%
      \put(9984,514){\makebox(0,0){\strut{}0.01}}%
      \put(10979,514){\makebox(0,0){\strut{}0.014}}%
    }%
    \gplbacktext
    \put(0,0){\includegraphics{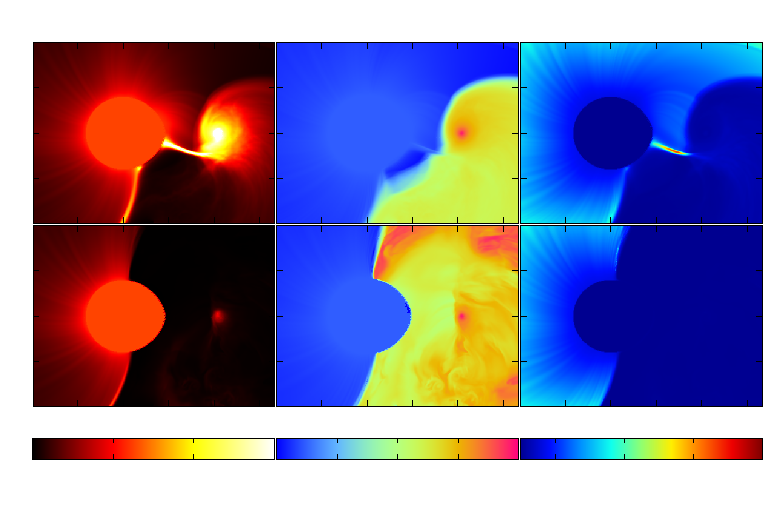}}%
    \gplfronttext
  \end{picture}%
\endgroup

%% file: TP_12.tex
\begingroup
  \makeatletter
  \providecommand\color[2][]{%
    \GenericError{(gnuplot) \space\space\space\@spaces}{%
      Package color not loaded in conjunction with
      terminal option `colourtext'%
    }{See the gnuplot documentation for explanation.%
    }{Either use 'blacktext' in gnuplot or load the package
      color.sty in LaTeX.}%
    \renewcommand\color[2][]{}%
  }%
  \providecommand\includegraphics[2][]{%
    \GenericError{(gnuplot) \space\space\space\@spaces}{%
      Package graphicx or graphics not loaded%
    }{See the gnuplot documentation for explanation.%
    }{The gnuplot epslatex terminal needs graphicx.sty or graphics.sty.}%
    \renewcommand\includegraphics[2][]{}%
  }%
  \providecommand\rotatebox[2]{#2}%
  \@ifundefined{ifGPcolor}{%
    \newif\ifGPcolor
    \GPcolortrue
  }{}%
  \@ifundefined{ifGPblacktext}{%
    \newif\ifGPblacktext
    \GPblacktextfalse
  }{}%
  \let\gplgaddtomacro\g@addto@macro
  \gdef\gplbacktext{}%
  \gdef\gplfronttext{}%
  \makeatother
  \ifGPblacktext
    \def\colorrgb#1{}%
    \def\colorgray#1{}%
  \else
    \ifGPcolor
      \def\colorrgb#1{\color[rgb]{#1}}%
      \def\colorgray#1{\color[gray]{#1}}%
      \expandafter\def\csname LTw\endcsname{\color{white}}%
      \expandafter\def\csname LTb\endcsname{\color{black}}%
      \expandafter\def\csname LTa\endcsname{\color{black}}%
      \expandafter\def\csname LT0\endcsname{\color[rgb]{1,0,0}}%
      \expandafter\def\csname LT1\endcsname{\color[rgb]{0,1,0}}%
      \expandafter\def\csname LT2\endcsname{\color[rgb]{0,0,1}}%
      \expandafter\def\csname LT3\endcsname{\color[rgb]{1,0,1}}%
      \expandafter\def\csname LT4\endcsname{\color[rgb]{0,1,1}}%
      \expandafter\def\csname LT5\endcsname{\color[rgb]{1,1,0}}%
      \expandafter\def\csname LT6\endcsname{\color[rgb]{0,0,0}}%
      \expandafter\def\csname LT7\endcsname{\color[rgb]{1,0.3,0}}%
      \expandafter\def\csname LT8\endcsname{\color[rgb]{0.5,0.5,0.5}}%
    \else
      \def\colorrgb#1{\color{black}}%
      \def\colorgray#1{\color[gray]{#1}}%
      \expandafter\def\csname LTw\endcsname{\color{white}}%
      \expandafter\def\csname LTb\endcsname{\color{black}}%
      \expandafter\def\csname LTa\endcsname{\color{black}}%
      \expandafter\def\csname LT0\endcsname{\color{black}}%
      \expandafter\def\csname LT1\endcsname{\color{black}}%
      \expandafter\def\csname LT2\endcsname{\color{black}}%
      \expandafter\def\csname LT3\endcsname{\color{black}}%
      \expandafter\def\csname LT4\endcsname{\color{black}}%
      \expandafter\def\csname LT5\endcsname{\color{black}}%
      \expandafter\def\csname LT6\endcsname{\color{black}}%
      \expandafter\def\csname LT7\endcsname{\color{black}}%
      \expandafter\def\csname LT8\endcsname{\color{black}}%
    \fi
  \fi
  \setlength{\unitlength}{0.0500bp}%
  \begin{picture}(12960.00,5328.00)%
    \gplgaddtomacro\gplbacktext{%
    }%
    \gplgaddtomacro\gplfronttext{%
      \csname LTb\endcsname%
      \put(1706,710){\makebox(0,0){\strut{}-600}}%
      \put(2760,710){\makebox(0,0){\strut{}-300}}%
      \put(3859,710){\makebox(0,0){\strut{}0}}%
      \put(4956,710){\makebox(0,0){\strut{}300}}%
      \put(6054,710){\makebox(0,0){\strut{}600}}%
      \put(1535,996){\makebox(0,0)[r]{\strut{}-600}}%
      \put(1535,2039){\makebox(0,0)[r]{\strut{}-300}}%
      \put(1535,3127){\makebox(0,0)[r]{\strut{}0}}%
      \put(1535,4213){\makebox(0,0)[r]{\strut{}300}}%
      \put(1535,5300){\makebox(0,0)[r]{\strut{}600}}%
      \put(1711,0){\makebox(0,0){\strut{}0}}%
      \put(2781,0){\makebox(0,0){\strut{}0.25}}%
      \put(3851,0){\makebox(0,0){\strut{}0.50}}%
      \put(4921,0){\makebox(0,0){\strut{}0.75}}%
      \put(6013,0){\makebox(0,0){\strut{}1}}%
      \put(3244,5518){\makebox(0,0)[l]{\strut{}$\mbox{Low/Hard state}$}}%
      \put(4297,691){\makebox(0,0)[l]{\strut{}$v_\mathrm{x}$}}%
      \put(1399,3561){\rotatebox{-270}{\makebox(0,0)[l]{\strut{}$v_\mathrm{y}$}}}%
    }%
    \gplgaddtomacro\gplbacktext{%
    }%
    \gplgaddtomacro\gplfronttext{%
      \csname LTb\endcsname%
      \put(7783,699){\makebox(0,0){\strut{}-600}}%
      \put(8837,699){\makebox(0,0){\strut{}-300}}%
      \put(9936,699){\makebox(0,0){\strut{}0}}%
      \put(11033,699){\makebox(0,0){\strut{}300}}%
      \put(12131,699){\makebox(0,0){\strut{}600}}%
      \put(7612,985){\makebox(0,0)[r]{\strut{}-600}}%
      \put(7612,2028){\makebox(0,0)[r]{\strut{}-300}}%
      \put(7612,3116){\makebox(0,0)[r]{\strut{}0}}%
      \put(7612,4202){\makebox(0,0)[r]{\strut{}300}}%
      \put(7612,5289){\makebox(0,0)[r]{\strut{}600}}%
      \put(7801,0){\makebox(0,0){\strut{}0}}%
      \put(8871,0){\makebox(0,0){\strut{}0.25}}%
      \put(9941,0){\makebox(0,0){\strut{}0.50}}%
      \put(11011,0){\makebox(0,0){\strut{}0.75}}%
      \put(12103,0){\makebox(0,0){\strut{}1}}%
      \put(9321,5507){\makebox(0,0)[l]{\strut{}$\mbox{High/Soft state}$}}%
      \put(10374,680){\makebox(0,0)[l]{\strut{}$v_\mathrm{x}$}}%
      \put(7476,3550){\rotatebox{-270}{\makebox(0,0)[l]{\strut{}$v_\mathrm{y}$}}}%
    }%
    \gplbacktext
    \put(0,0){\includegraphics{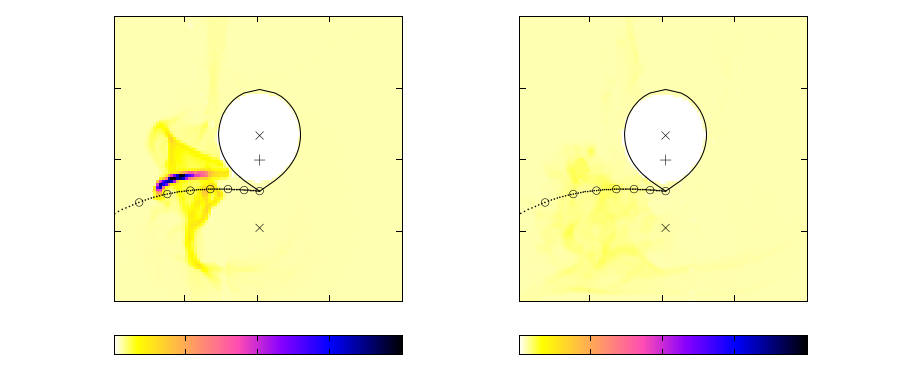}}%
    \gplfronttext
  \end{picture}%
\endgroup

%% file: TP_13.tex
\begingroup
  \makeatletter
  \providecommand\color[2][]{%
    \GenericError{(gnuplot) \space\space\space\@spaces}{%
      Package color not loaded in conjunction with
      terminal option `colourtext'%
    }{See the gnuplot documentation for explanation.%
    }{Either use 'blacktext' in gnuplot or load the package
      color.sty in LaTeX.}%
    \renewcommand\color[2][]{}%
  }%
  \providecommand\includegraphics[2][]{%
    \GenericError{(gnuplot) \space\space\space\@spaces}{%
      Package graphicx or graphics not loaded%
    }{See the gnuplot documentation for explanation.%
    }{The gnuplot epslatex terminal needs graphicx.sty or graphics.sty.}%
    \renewcommand\includegraphics[2][]{}%
  }%
  \providecommand\rotatebox[2]{#2}%
  \@ifundefined{ifGPcolor}{%
    \newif\ifGPcolor
    \GPcolortrue
  }{}%
  \@ifundefined{ifGPblacktext}{%
    \newif\ifGPblacktext
    \GPblacktextfalse
  }{}%
  \let\gplgaddtomacro\g@addto@macro
  \gdef\gplbacktext{}%
  \gdef\gplfronttext{}%
  \makeatother
  \ifGPblacktext
    \def\colorrgb#1{}%
    \def\colorgray#1{}%
  \else
    \ifGPcolor
      \def\colorrgb#1{\color[rgb]{#1}}%
      \def\colorgray#1{\color[gray]{#1}}%
      \expandafter\def\csname LTw\endcsname{\color{white}}%
      \expandafter\def\csname LTb\endcsname{\color{black}}%
      \expandafter\def\csname LTa\endcsname{\color{black}}%
      \expandafter\def\csname LT0\endcsname{\color[rgb]{1,0,0}}%
      \expandafter\def\csname LT1\endcsname{\color[rgb]{0,1,0}}%
      \expandafter\def\csname LT2\endcsname{\color[rgb]{0,0,1}}%
      \expandafter\def\csname LT3\endcsname{\color[rgb]{1,0,1}}%
      \expandafter\def\csname LT4\endcsname{\color[rgb]{0,1,1}}%
      \expandafter\def\csname LT5\endcsname{\color[rgb]{1,1,0}}%
      \expandafter\def\csname LT6\endcsname{\color[rgb]{0,0,0}}%
      \expandafter\def\csname LT7\endcsname{\color[rgb]{1,0.3,0}}%
      \expandafter\def\csname LT8\endcsname{\color[rgb]{0.5,0.5,0.5}}%
    \else
      \def\colorrgb#1{\color{black}}%
      \def\colorgray#1{\color[gray]{#1}}%
      \expandafter\def\csname LTw\endcsname{\color{white}}%
      \expandafter\def\csname LTb\endcsname{\color{black}}%
      \expandafter\def\csname LTa\endcsname{\color{black}}%
      \expandafter\def\csname LT0\endcsname{\color{black}}%
      \expandafter\def\csname LT1\endcsname{\color{black}}%
      \expandafter\def\csname LT2\endcsname{\color{black}}%
      \expandafter\def\csname LT3\endcsname{\color{black}}%
      \expandafter\def\csname LT4\endcsname{\color{black}}%
      \expandafter\def\csname LT5\endcsname{\color{black}}%
      \expandafter\def\csname LT6\endcsname{\color{black}}%
      \expandafter\def\csname LT7\endcsname{\color{black}}%
      \expandafter\def\csname LT8\endcsname{\color{black}}%
    \fi
  \fi
  \setlength{\unitlength}{0.0500bp}%
  \begin{picture}(12960.00,5328.00)%
    \gplgaddtomacro\gplbacktext{%
    }%
    \gplgaddtomacro\gplfronttext{%
      \csname LTb\endcsname%
      \put(985,691){\makebox(0,0){\strut{}-1}}%
      \put(2001,691){\makebox(0,0){\strut{}-0.5}}%
      \put(3031,691){\makebox(0,0){\strut{}0}}%
      \put(4060,691){\makebox(0,0){\strut{}0.5}}%
      \put(5090,691){\makebox(0,0){\strut{}1}}%
      \put(6120,691){\makebox(0,0){\strut{}1.5}}%
      \put(813,977){\makebox(0,0)[r]{\strut{}-1}}%
      \put(813,1994){\makebox(0,0)[r]{\strut{}-0.5}}%
      \put(813,3024){\makebox(0,0)[r]{\strut{}0}}%
      \put(813,4053){\makebox(0,0)[r]{\strut{}0.5}}%
      \put(813,5083){\makebox(0,0)[r]{\strut{}1}}%
      \put(1768,0){\makebox(0,0){\strut{}0.002}}%
      \put(3335,0){\makebox(0,0){\strut{}0.006}}%
      \put(4902,0){\makebox(0,0){\strut{}0.01}}%
      \put(6469,0){\makebox(0,0){\strut{}0.014}}%
      \put(3600,5303){\makebox(0,0)[l]{\strut{}$n_\mathrm{H}$}}%
      \colorrgb{1.00,1.00,1.00}%
      \put(4142,2708){\makebox(0,0)[l]{\strut{}$1$}}%
      \put(4458,2625){\makebox(0,0)[l]{\strut{}$2$}}%
      \put(4142,3146){\makebox(0,0)[l]{\strut{}$3$}}%
      \put(4513,3146){\makebox(0,0)[l]{\strut{}$4$}}%
      \put(3964,3558){\makebox(0,0)[l]{\strut{}$5$}}%
      \put(4582,3558){\makebox(0,0)[l]{\strut{}$6$}}%
      \put(4747,3970){\makebox(0,0)[l]{\strut{}$7$}}%
      \put(4994,2461){\makebox(0,0)[l]{\strut{}$8$}}%
      \put(5749,2461){\makebox(0,0)[l]{\strut{}$9$}}%
      \put(5406,2048){\makebox(0,0)[l]{\strut{}$10$}}%
    }%
    \gplgaddtomacro\gplbacktext{%
    }%
    \gplgaddtomacro\gplfronttext{%
      \csname LTb\endcsname%
      \put(7871,691){\makebox(0,0){\strut{}-600}}%
      \put(8876,691){\makebox(0,0){\strut{}-300}}%
      \put(9924,691){\makebox(0,0){\strut{}0}}%
      \put(10970,691){\makebox(0,0){\strut{}300}}%
      \put(12017,691){\makebox(0,0){\strut{}600}}%
      \put(7699,977){\makebox(0,0)[r]{\strut{}-600}}%
      \put(7699,1973){\makebox(0,0)[r]{\strut{}-300}}%
      \put(7699,3010){\makebox(0,0)[r]{\strut{}0}}%
      \put(7699,4046){\makebox(0,0)[r]{\strut{}300}}%
      \put(7699,5083){\makebox(0,0)[r]{\strut{}600}}%
      \put(7866,0){\makebox(0,0){\strut{}0}}%
      \put(8897,0){\makebox(0,0){\strut{}0.25}}%
      \put(9929,0){\makebox(0,0){\strut{}0.50}}%
      \put(10961,0){\makebox(0,0){\strut{}0.75}}%
      \put(12013,0){\makebox(0,0){\strut{}1}}%
      \put(9882,5290){\makebox(0,0)[l]{\strut{}$I$}}%
      \put(10341,687){\makebox(0,0)[l]{\strut{}$v_\mathrm{x}$}}%
      \put(7578,3424){\rotatebox{-270}{\makebox(0,0)[l]{\strut{}$v_\mathrm{y}$}}}%
      \colorrgb{1.00,1.00,1.00}%
      \put(9115,2238){\makebox(0,0)[l]{\strut{}$10$}}%
      \put(8839,2284){\makebox(0,0)[l]{\strut{}$8$}}%
      \put(8964,1865){\makebox(0,0)[l]{\strut{}$9$}}%
      \put(8755,2699){\makebox(0,0)[l]{\strut{}$2$}}%
      \put(9216,2740){\makebox(0,0)[l]{\strut{}$1$}}%
      \put(9216,3030){\makebox(0,0)[l]{\strut{}$3$}}%
      \put(8663,3361){\makebox(0,0)[l]{\strut{}$6$}}%
      \put(9052,2574){\makebox(0,0)[l]{\strut{}$5$}}%
      \put(8717,3278){\makebox(0,0)[l]{\strut{}$4$}}%
      \put(8717,2948){\makebox(0,0)[l]{\strut{}$7$}}%
    }%
    \gplbacktext
    \put(0,0){\includegraphics{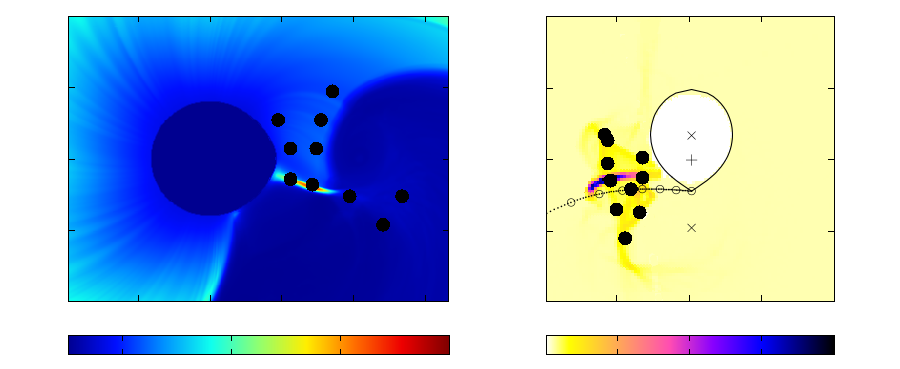}}%
    \gplfronttext
  \end{picture}%
\endgroup